\documentclass[reprint, superscriptaddress, nofootinbib, amsmath, amssymb, aps, pra
]{revtex4-2}

\usepackage{graphicx}
\usepackage{dcolumn}
\usepackage{bm}
\usepackage{hyperref}
\hypersetup{
    colorlinks=true,
    linkcolor=blue,
    filecolor=blue,      
    urlcolor=blue,
    citecolor=blue, ,
    pdftitle={PTA Per-frequency Optimal Statistic},
    pdfpagemode=FullScreen,
    }
\usepackage[mathlines]{lineno}
\usepackage{appendix}
\usepackage{cancel}
\usepackage{aas_macros}

\newcommand{\trace}[1]{\mathrm{tr}\left [ #1 \right ]}
\newcommand{\expect}[1]{\left < #1 \right >}
\newcommand{\blockComment}[1]{}

\begin{document}

\title{Spatial and Spectral Characterization of the Gravitational-wave Background \\ with the PTA Optimal Statistic}

\author{Kyle A. Gersbach}
\email{kyle.a.gersbach@vanderbilt.edu}
\affiliation{Department of Physics \& Astronomy, Vanderbilt University,\\
2301 Vanderbilt Place, Nashville, TN 37235, USA}

\author{Stephen R. Taylor}
\affiliation{Department of Physics \& Astronomy, Vanderbilt University,\\
2301 Vanderbilt Place, Nashville, TN 37235, USA}

\author{Patrick M. Meyers}
\affiliation{Division of Physics, Mathematics, and Astronomy, California Institute of Technology,\\ 
Pasadena, CA 91125, USA}

\author{Joseph D. Romano}
\affiliation{Department of Physics and Astronomy, University of Texas Rio Grande Valley,\\
One West University Boulevard, Brownsville, TX 78520, USA}

\date{\today}

\begin{abstract}
Pulsar timing arrays (PTAs) have made tremendous progress and are now showing strong evidence for the gravitational-wave background (GWB). Further probing the origin and characteristics of the GWB will require more generalized analysis techniques. Bayesian methods are most often used but can be computationally expensive. On the other hand, frequentist methods, like the PTA Optimal Statistic (OS), are more computationally efficient and can produce results that are complementary to Bayesian methods, allowing for stronger statistical cases to be built from a confluence of different approaches. In this work we expand the capabilities of the OS through a technique we call the Per-Frequency Optimal Statistic (PFOS). The PFOS removes the underlying power-law assumption inherent in previous implementations of the OS, and allows one to estimate the GWB spectrum in a frequency-by-frequency manner. We have also adapted a recent generalization from the OS pipeline into the PFOS, making it capable of accurately characterizing the spectrum in the intermediate and strong GW signal regimes using only a small fraction of the necessary computational resources when compared with fully-correlated Bayesian methods, while also empowering many new types of analyses not possible before. We find that even in the strong GW signal regime, where the GWB dominates over noise in all frequencies, the injected value of the signal lies within the $50^\mathrm{th}$-percentile of the PFOS uncertainty distribution in $41-45\%$ of simulations, remaining $3\sigma$-consistent with unbiased estimation.
\end{abstract}


\maketitle

\section{Introduction} \label{sec:intro}

Pulsar timing arrays (PTAs) \citep{foster1990PTA} exploit the precision timing of millisecond-period pulsars to search for interpulsar-correlated timing residuals over campaigns of years to decades. This timespan sets the period (and frequencies, i.e., nanohertz) of GWs to which PTAs are sensitive. At such low frequencies and with initially limited resolution, PTAs are primarily sensitive to a GW background (GWB) \citep{rosado2015expected, kelley2018single}. If this background is isotropic then the timing correlation signature is called the Hellings \& Downs (HD) curve \citep{Hellings1983}, a quadrupolar-like function of only the angular separation of pulsars. In June 2023, several major PTA collaborations announced evidence for this GWB, including the North American Nanohertz Observatory for Gravitational waves (NANOGrav) with $\sim4\sigma$ significance \citep{NG15_evidence}, the European and Indian collaborations (EPTA and InPTA) with joint $\sim3\sigma$ significance \citep{Antoniadis2023}, and the Parkes Pulsar Timing Array (PPTA) with $\sim2\sigma$ significance \citep{Reardon2023}. The International Pulsar Timing Array (IPTA) conducted a comparison of all results, which validated their consistency in terms of relative detection sensitivity and their measurements of the GWB spectrum \citep{Zic2023}.
\footnote{The more recently established Chinese PTA also announced evidence, albeit only at one frequency and using statistical techniques that are not comparable with those used by the other PTAs \citep{CPTA}.} 

As datasets grow in sensitivity to the GWB, its spatial and spectral properties will become ever more of interest in the goal of understanding its origin \citep{pol2022anisotropy, Lamb2023, taylor2020bumpy}. The primary signal hypothesis is of a large population of supermassive black-hole binary (SMBHB) systems ($\sim10^8-10^{10}M_\odot$) at sub-centiparsec separations, all emitting GWs that sum incoherently to produce a stochastic timing signal \citep{Phinney2001, NG15_evidence, Antoniadis2023}. Other interpretations posit that part or all of the GWB is from inflationary or early-Universe phase-transition processes, topological defects, or dark matter \citep{NG15_AstroInterp, wu2024cosmological}. While many current inference techniques assume power-law models for the GWB characteristic strain spectrum, the effects of discreteness \citep[e.g.,][]{becsy2022realistic, NG15_discret} or non-GW--driven orbital evolution in a SMBHB population \citep{sampson2015constraining,taylor2017constraints}, or the details of cosmological processes \citep{NG15_newphysics, kibble1976topology}, can all induce departures from such simple behavior. There is a growing need for detection statistics and parameter estimation techniques that can generalize to arbitrary spectral and spatial power distributions, and do so rapidly for large-scale programs of simulations and explorations.

The most commonly used framework in PTA searches is that of Bayesian inference, whereupon a generative model of pulsar timing residuals across an array is constructed. This model is then assessed against data in a likelihood function, and with parameter prior assumptions, the joint posterior probability distribution of model parameters can be sampled using Markov Chain Monte Carlo (MCMC) techniques \citep{Johnson2023, Taylor2021}. One can then construct marginals of this joint distribution to derive parameter estimates and uncertainties. Using a variety of approaches, ratios of the fully marginalized likelihood (i.e., Bayes factors) can also be calculated to act as detection statistics; in PTAs the relevant ratio is of a steep-spectrum red process that is either HD-correlated or uncorrelated between pulsars. The flexibility of Bayesian modeling makes it the primary statistical approach, yet its major downside is computational expense, with the bottleneck of the likelihood function being covariance-matrix inversions that must be performed sequentially in MCMC analyses that are run for millions of steps. Despite this, there are new methods that are accelerating these steps \citep{taylor2022parallelized, Lamb2023, freedman2023efficient, hourihane2023accurate}. 

An alternative approach, and one which is most often used for determining detection significance, is the PTA optimal statistic (OS) \citep{Anholm2009}. This is a frequentist statistic originally developed to look for GWBs in ground-based gravitational-wave detectors \citep{Allen1999}. In a PTA context, the OS was conceived to compute the amplitude and signal-to-noise ratio (S/N) of a GWB with a power-law strain spectrum, usually fixed to the expected shape given by a SMBHB population \citep{Anholm2009, Chamberlin2015}. 
While other frequentist GWB detection techniques also exist and have been used in some recent publications (i.e., \citet{Jenet_2005} was used in \citet{CPTA}), the OS was shown to be both a minimum variance estimator within noise dominated data and an optimal detection statistic \citep{Anholm2009}.
Unlike the Bayesian approach, the OS focuses exclusively on the cross-correlated signal between pulsars, leveraging the distinctiveness of the HD curve and treating the auto-correlated GWB signal as a noise process that adds to intrinsic noise in the pulsars. It is also a much faster statistic to evaluate, which means that it is often the first approach carried out to provide detection statistics, GWB amplitude estimates, reconstructed correlation signatures, and cross-validation tests. Its speed also makes calibration of the OS S/N with large numbers of data augmentation operations (e.g., phase shifting \cite{Taylor2017} and sky scrambling \cite{cornish2016towards}) more tractable.

Much work has recently been done to extend the OS's applicability and usefulness. It was originally developed under the weak-signal hypothesis, where the GWB signal would always be buried in noise. However, this is no longer true of the PTA GWB signal \citep{Romano2021}, meaning that the assumptions of the OS required revision to account for pairwise correlation measurements of the GWB no longer being independent \citep{allen-romano2023,Allen2023}. Furthermore, while the OS was originally designed to search for just an isotropic GWB, new work has extended it to search for anisotropy \citep{pol2022anisotropy}, as well as multiple processes among which may be systematics that induce spatial correlations (e.g., clock errors or Solar System ephemeris offsets) \citep{Sardesai2023}. Some of these recent advances were used in reporting NANOGrav's evidence for the GWB \citep{NG15_evidence,Vallisneri2023,Meyers2023}. These new methods and extensions, along with the ones we propose in this paper, are briefly explained in \autoref{tab:os_extensions}. Importantly, all of these methods can be combined together to form even more powerful methods.

\begin{table*}[t!]
\begin{tabular}{|c|c|l|c|c|}
\hline
\textbf{Method Name} & \textbf{Abbreviation} & \multicolumn{1}{c|}{\textbf{Description}} & \textbf{Section} & \textbf{Citation} \\ 
\hline

Optimal Statistic & 
OS & 
\begin{tabular}[c]{@{}l@{}}
    The first PTA optimal statistic. Tests the significance of a \\ 
    single spatial correlation pattern. Biased in intermediate \\ 
    and strong signal regimes.
\end{tabular} &
\ref{sec:OS} &
\citep{Anholm2009,Chamberlin2015} \\ 
\hline

Noise-Marginalized OS & 
NMOS & 
\begin{tabular}[c]{@{}l@{}}
    An additional procedure to account for the spread in \\ 
    parameter estimates from a Bayesian 
    common uncorrelated \\
    red noise search by iteratively running the OS with many \\ 
    parameter estimates. More accurate in cases with significant \\
    intrinsic red noise. 
\end{tabular} &
\ref{sec:unc_samp} &
\citep{Vigeland2018} \\ 
\hline

Pair-Covariant OS & 
PCOS & 
\begin{tabular}[c]{@{}l@{}}
    An additional procedure to incorporate covariance between\\ 
    pulsar pairs in cases with a significant GWB. Fixes OS bias \\ 
    in intermediate and strong signal regimes.
\end{tabular} &
\ref{sec:par_est} &
\citep{Johnson2023,Romano2021,allen-romano2023} \\ 
\hline

Multiple-Component OS & 
MCOS & 
\begin{tabular}[c]{@{}l@{}}
    A generalization of the OS which allows for simultaneous \\ 
    estimates of multiple spatial correlation pattern amplitudes.
\end{tabular} &
-- &
\citep{Sardesai2023} \\ 
\hline

Per-Frequency OS* &
PFOS & 
\begin{tabular}[c]{@{}l@{}}
    A generalization to the OS which allows for estimating a \\
    single correlation pattern at individual frequencies. \\
    Accounts for GWB amplitude at other frequencies.
\end{tabular} &
\ref{sec:PFOS_derivation} &
-- \\ 
\hline

\begin{tabular}[c]{@{}c@{}}Narrowband-normalized \\ Per-Frequency OS*\end{tabular} & 
\begin{tabular}[c]{@{}c@{}}Narrowband\\ PFOS\end{tabular} &
\begin{tabular}[c]{@{}l@{}}
    A simplification of the PFOS which estimates the GWB \\ 
    amplitude at each frequency assuming other frequencies \\ 
    don't contribute. Biased in strong signal regimes.
\end{tabular} &
\ref{sec:PFOS_derivation} &
-- \\ 
\hline

Uncertainty Sampling* & 
UC & 
\begin{tabular}[c]{@{}l@{}}
    A method that more accurately describes the total GWB \\ 
    amplitude from a NMOS analysis by accounting for the \\ 
    uncertainties from each OS iteration.
\end{tabular} &
\ref{sec:unc_samp} &
-- \\ 
\hline

\end{tabular}

\caption{A table listing the many methods and extensions of the PTA optimal statistic. These methods can also be combined together to form a compound method such as the Pair-Covariant Noise-Marginalized Optimal Statistic (PC+NMOS). Names marked with a `*' indicate methods which are new in this work.}

\label{tab:os_extensions}
\end{table*}

However, a core assumption of the OS still remains that the GWB has a power-law strain spectrum, often kept at the fiducial value for a SMBHB population. In this work, we generalize the OS beyond this assumption, allowing for our new per-frequency optimal statistic (PFOS) to characterize the GWB spectrum and its spatial-correlation properties using frequentist statistics. Our method not only provides a more general frequentist framework for PTA GWB searches, but also allows for comparisons with Bayesian measurements to strengthen future GWB evidence, and potentially even show early signs of an emerging individual GW source. 

This paper is laid out as follows. In \autoref{sec:OS} we review the PTA optimal statistic as it is currently implemented, as well as its recent generalization to the arbitrary signal-strength regime. \autoref{sec:PFOS_derivation} introduces our new PTA per-frequency statistic, allowing for generalized spectral estimation that can nevertheless be subsequently fit to parameterized spectral models (e.g., power laws, models with low-frequency turnovers, or models with single-frequency excess due to discrete binary population behavior). We discuss the design of simulations with which to assess the efficacy of this new statistic in \autoref{sec:sim_design} followed by the results of these assessments in \autoref{sec:method_test}. Our conclusions and discussion of future work are presented in \autoref{sec:discussion}, followed by several appendices of supporting calculations.

\section{Traditional Optimal statistic} \label{sec:OS}

The optimal statistic (OS) is a cross-correlation statistic employed in GWB searches with PTA experiments \citep{Anholm2009, Chamberlin2015, Siemens2013}; its name comes from the use of an optimal filter in the cross-correlation calculation that maximizes the signal-to-noise ratio \citep{Allen1999, Anholm2009}. It is primarily used as a frequentist detection statistic since it can be quickly evaluated, and is often employed before or in parallel with more expensive Bayesian analyses. To explain these statements in more detail, this section will focus on the OS as originally derived (henceforth called the \textit{traditional OS} for clarity) and discuss its primary use cases. A reader familiar with the traditional OS may wish to skip to \autoref{sec:PFOS_derivation} where we introduce our new techniques.

\subsection{Creating the Traditional OS }
\label{sec:preliminaries}

The measurements made by PTAs come in the form of vectors of pulse times of arrival (TOAs) for all pulsars in our PTA, $\bm{t}_{\mathrm{TOA}}$. While many possible processes can change the expected arrival time of these pulses, each process can be broadly grouped into either deterministic or stochastic processes \citep{Taylor2021}. The leading-order deterministic contribution is the timing model, whose parameters are estimated through individual pulsar analyses and are used to predict the expected TOAs, $\bm{t}_{\mathrm{det}}$ \citep{Taylor2021}. By subtracting the best-fit timing model from the measured TOAs, we can create a set of timing residuals,
\begin{equation}
    \bm{\delta t} = \bm{t}_{\mathrm{TOA}} - \bm{t}_{\mathrm{det}}
    .
\end{equation}
 
When modeling these residuals, we must also consider stochastic processes, as well as allowing for linearized offsets in the deterministic model fits. Hence we model these residuals as
\begin{equation}
    \bm{\delta t} = \bm{M} \vec{\epsilon} + \bm{F} \vec{a} + \vec{n}_{w},
    \label{eq:residuals}
\end{equation}
where the linearized timing model design matrix, $\bm{M}$, contains partial derivatives of TOAs with respect to each timing model parameter, allowing for deviations from the best-fit deterministic model found in individual pulsar analyses; $\vec{\epsilon}$ are linear timing model parameter deviations; $\bm{F}$ is the total red-noise Fourier design matrix which contains alternating TOA-length columns of sine and cosine elements evaluated at each of the lowest frequencies, modeling both intrinsic red noise in a single pulsar as well as the common red noise from the GWB; $\vec{a}$ are the corresponding Fourier coefficients; and $\vec{n}_\mathrm{w}$ is an additional white noise term intrinsic to each pulsar and  mostly accounted for by TOA measurement uncertainties.

We can write a Gaussian likelihood function over all residuals in the PTA,
\begin{equation}
    p(\bm{\delta t}|\vec{\epsilon},\vec{a}) = 
    \frac{1}{\sqrt{\mathrm{det}(2 \pi \bm{\Sigma})}} \exp{\left(-\frac{1}{2}\bm{\delta t}^T \bm{\Sigma}^{-1} \bm{\delta t}\right)}
    .
    \label{eq:likelihood}
\end{equation}
The residual covariance matrix, $\bm{\Sigma}$, is modeled through blocks of individual pulsar auto- and cross-covariances,
\begin{equation}
    \bm{\Sigma} = 
        \begin{bmatrix}
        \bm{P}_1 & \bm{S}_{12} & \bm{S}_{13} & \\
        \bm{S}_{21} & \bm{P}_{2} & \bm{S}_{23} & ...\\
        \bm{S}_{31} & \bm{S}_{32} & \bm{P}_{3} & \\
        & \vdots & 
        \end{bmatrix}
    ,
    \label{eq:SigMatrix}
\end{equation}
where we define sub-matrices for auto-covariance, $\bm{P}_a = \expect{\bm{\delta t}_a \bm{\delta t}_a^T}$, and cross-covariances, $\bm{S}_{ab} = \expect{\bm{\delta t}_a \bm{\delta t}_b^T}$ for $a \ne b$. In these equations, the subscripts $a$ and $b$ label pulsars, such that, e.g., $\bm{\delta t}_a$ is a column vector of timing residuals for pulsar $a$. The GWB induces cross-covariances in the timing residuals of pulsars, such that
\begin{equation}
    \bm{S}_{ab} = \Gamma_{ab} \; \bm{F}_a \phi \bm{F}_b^T,
    \label{eq:correlation_sub}
\end{equation}
where $\Gamma_{ab}$ represents the constant factor of the overlap reduction function (ORF) between pulsars $a$ and $b$, $\bm{F}_a$ is pulsar $a$'s Fourier design matrix, and $\phi$ is the diagonal Fourier-domain covariance matrix of timing residuals induced by the GWB, represented here as the power spectral density (PSD) of the GWB in the sine and cosine elements of each Fourier frequency. These frequencies are modeled as multiples of a base Fourier frequency, $f_k = k / T_{\mathrm{span}}$, where $T_{\mathrm{span}}$ is the total time-span of the PTA, $k\in[1,2,\ldots,k_\mathrm{max}]$, and $k_\mathrm{max}\ll n_\mathrm{TOA}$. 

The traditional OS assumes that the GWB's PSD, $S(f_k)$, is described by a power-law, such that for the different frequencies, $f_k$, of our PTA we can write the diagonal elements of the Fourier domain covariance matrix as
\begin{equation}
    \phi_{kk'} = \delta_{kk'} S(f_k) \Delta f = 
    \frac{A_{\mathrm{gw}}^2 }{12 \pi^2 T_{\mathrm{span}}}
    \left( \frac{f_k}{f_{\mathrm{yr}}} \right)
    ^{-\gamma} f_{\mathrm{yr}}^{-3}
    ,
    \label{eq:powerlaw}
\end{equation}
where $\Delta f = T_{\mathrm{span}}^{-1}$ is the width of a frequency bin, $A_{\mathrm{gw}}$ is the amplitude of the GWB characteristic strain at a reference frequency of $f_{\mathrm{yr}}=1/\mathrm{year}$, and $\gamma$ is the spectral index of the power-law, which is expected to be $13/3$ for a GWB composed of SMBHBs evolving entirely due to gravitational waves \citep{Phinney2001}. Note that the off-diagonals of this covariance matrix are zero since we model the GWB and all red processes as stationary. The shape of the $\phi$ matrix is ($2k_\mathrm{max}\times 2k_\mathrm{max}$) since $\phi$ includes both the sine and cosine variances of each frequency. Also note that for ease of notation, we suppress the $\Delta f$ factor from the PSD for the remainder of the paper, treating the PSD to be synonymous with variance of timing residuals at each frequency.

The ORF can be generalized to be any number of other correlation patterns like monopole, dipole, or even a combination of those correlation patterns ~\citep{Sardesai2023}. For simplicity, this paper uses only Hellings and Downs (HD) correlations, as that is what is expected for an isotropic GWB \citep{Hellings1983}. The HD ORF can be written as,
\begin{equation}
    \Gamma_{ab} = \frac{1}{2} 
    - \frac{1-\cos{\xi_{ab}}}{4} 
    \left[ \frac{1}{2} - 3 \ln \left( \frac{1-\cos{\xi_{ab}}}{2} \right) \right] + \frac{1}{2}\delta_{ab}
    ,
    \label{eq:hd_orf}
\end{equation}
where $\xi_{ab}$ is the separation angle between pulsars $a$ and $b$, and $\delta_{ab}$ is a Kronecker delta function that accounts for the pulsar term being fully correlated only when $a=b$.

With these now defined, we can introduce the traditional OS. \citet{Chamberlin2015} show that by making a first-order Taylor expansion of the PTA log-likelihood, the traditional OS can be derived as a maximum-likelihood estimator of $A^2_{\mathrm{gw}}$. Importantly, this derivation assumes that the PTA is in the weak-signal regime, meaning that the GWB is weaker than the sum of all other noise sources at each frequency analyzed. This assumption will be touched on later in \autoref{sec:par_est}. For this paper, we opt to use the version presented in Appendix A of \citet{Chamberlin2015}, which presents the traditional OS in terms of pair-wise estimators,
\begin{equation}
    \rho_{ab} =
    \mathcal{N}_{ab} \,
    \bm{\delta t}_a^T \bm{P}_a^{-1} \hat{\bm{S}}_{ab} \bm{P}_b^{-1} \bm{\delta t}_b
    .
    \label{eq:os_pre_norm}
\end{equation}

In this equation, $\bm{\delta t}_a$ and $\bm{P}_a$ are as previously defined, and $\hat{\bm{S}}_{ab}$ is the cross-covariance matrix between $a$ and $b$ divided by the ORF factor and amplitude of the GWB power-law, such that $\bm{S}_{ab} = A^2_{\mathrm{gw}} \Gamma_{ab} \hat{\bm{S}}_{ab}$. The normalization, $\mathcal{N}_{ab}$, is chosen such that the expectation is the ORF-modulated power-law GWB amplitude,
\begin{equation}
    \expect{\rho_{ab}} = \Gamma_{ab} A^2_{\mathrm{gw}}
    .
\end{equation}
This calculation is given in full in the Appendix \ref{sec:os_normalization}, and leads us to our final pair-wise estimator,
\begin{equation}
    \rho_{ab} =
    \frac 
    {\bm{\delta t}_a^T \bm{P}_a^{-1} \: \hat{\bm{S}}_{ab} \: \bm{P}_b^{-1} \bm{\delta t}_b}
    {\trace{ \bm{P}_b^{-1} \: \hat{\bm{S}}_{ba} \:
    \bm{P}_a^{-1} \: \hat{\bm{S}}_{ab} }}
    .
    \label{eq:rho_ij}
\end{equation}
The uncertainty in this estimator is derived by calculating its variance. As originally constructed and presented elsewhere, this quantity is derived under the null hypothesis of zero inter-pulsar correlations, i.e., $\expect{\rho_{ab}}_0 = 0$. Using this, and as derived in Appendix \ref{sec:os_sigma_calc}, the uncertainty on our pair-wise estimator is
\begin{equation}
    \sigma_{ab,0} =
    \trace{ \bm{P}_b^{-1} \: \hat{\bm{S}}_{ba} \: 
    \bm{P}_a^{-1} \: \hat{\bm{S}}_{ab} }^{-1/2},
    \label{eq:sigma_ij}
\end{equation}
where the subscript 0 indicates that this uncertainty has been derived under the null hypothesis.

Using the set of all pair-wise estimators we then can create the full traditional PTA optimal statistic through an inverse noise-weighted sum of the pair-wise estimators \citep{Chamberlin2015},
\begin{equation}
    \hat{A}^2_{\mathrm{gw}} = \frac
    {\sum_{a<b} \: \rho_{ab} \Gamma_{ab} / \sigma^2_{ab,0}}
    {\sum_{a<b} \: \Gamma^2_{ab} / \sigma^2_{ab,0}}
    .
    \label{eq:os_by_default}
\end{equation}
The associated uncertainty in this estimator is then calculated as the square root of the variance
\begin{equation}
    \sigma_{\hat{A}^2,0} = \left( \sum_{a<b} \frac{\Gamma_{ab}^2}{\sigma_{ab,0}^2} \right)^{-1/2} .
\end{equation}

Most of the matrices involved in these calculations are large and require correspondingly computationally expensive operations, such as the inverse matrix $\bm{P}^{-1}_{a}$ which has dimension equal to the squared number of TOAs for pulsar $a$. To help reduce the computational cost, we use rank-reduction strategies as detailed in Appendix A of \citet{pol2022anisotropy}. They define two matrix products,
\begin{equation}
    \bm{X}_a = \bm{F}_a^T \bm{P}_a^{-1} \bm{\delta t}_a
    ,
\end{equation}
\begin{equation}
    \bm{Z}_a = \bm{F}_a^T \bm{P}_a^{-1} \bm{F}_a
    .
    \label{eq:Z}
\end{equation}
Since these matrices now represent quantities in the Fourier domain, we will also need to decompose our residual cross-covariance matrices with the substitution, $\hat{\bm{S}}_{ab} = \bm{F}_a \hat{\phi} \bm{F}_b^T$, where $\phi = A^2_{\mathrm{gw}} \hat{\phi}$. Using these definitions, we can rewrite \autoref{eq:rho_ij} and \autoref{eq:sigma_ij} in a more computationally efficient form, 
\begin{equation}
    \rho_{ab} =
    \frac 
    {\bm{X}_a^T \: \hat{\phi} \: \bm{X}_b}
    {\trace{ \bm{Z}_b \: \hat{\phi} \:
    \bm{Z}_a \: \hat{\phi} }}
    ,
    \label{eq:rho_rank_reduced}
\end{equation}
\begin{equation}
    \sigma_{ab,0} =
    \trace{ \bm{Z}_b \: \hat{\phi} \: 
    \bm{Z}_a \: \hat{\phi}}^{-1/2}
    .
    \label{eq:sigma_rank_reduced}
\end{equation}

It is important to understand the many versions of $\phi$ that will appear in this paper. On its own $\phi$ represents the covariance matrix of the Fourier coefficients of the timing residuals induced by the GWB, which we take to be diagonal. $\hat{\phi}$ factors out the GWB amplitude making it a representation of the unit-amplitude power-law spectral template of the GWB.

To further generalize the traditional OS for future expansions, we frame \autoref{eq:os_by_default} as the solution to a chi-squared minimization scheme \citep{Chamberlin2015}, with 
\begin{equation}
    \chi^2 = 
    \left(\vec{\rho} - A^2_{\mathrm{gw}}\vec{\Gamma} \right)^T 
    \bm{C}_{0}^{-1}
    \left(\vec{\rho} - A^2_{\mathrm{gw}}\vec{\Gamma} \right)
    ,
\end{equation}
where we now represent all pulsar-pair estimators, $\rho_{ab}$, and corresponding ORF factors, $\Gamma_{ab}$, as column vectors of length equal to the number of distinct pulsar pairs in the PTA. $\bm{C}_{0}$ is the pulsar-pair covariance matrix assuming the null-hypothesis (indicated by the 0 subscript), whose elements are defined as
\begin{equation}
    \bm{C}_{0} = 
    \delta_{ac} \, \delta_{bd} \; \bm{C}_{ab,cd,0} =
    \delta_{ac} \, \delta_{bd} \; \sigma^2_{ab,0}.
    \label{eq:c_traditional}
\end{equation}
In the traditional OS derivation, this matrix is diagonal and trivial to invert. A more detailed discussion about this covariance matrix is given in \autoref{sec:par_est}.

Minimizing this chi-square results in a square amplitude estimator and a more generalizable form of \autoref{eq:os_by_default},
\begin{equation}
    \hat{A}^2_{\mathrm{gw}} = 
    \left( \vec{\Gamma}^T \bm{C}_{0}^{-1} \vec{\Gamma} \right)^{-1}
    \vec{\Gamma}^T \bm{C}_{0}^{-1} \vec{\rho}
    .
    \label{eq:Ahat}
\end{equation}
We then compute the uncertainty in this estimator under the null hypothesis by taking the square root of its variance,
\begin{equation}
    \sigma_{\hat{A}^2,0} =
    \left( \vec{\Gamma}^T \bm{C}_{0}^{-1} \vec{\Gamma} \right)^{-1/2}
    .
    \label{eq:sigma_ahat}
\end{equation}

Since our method modifies the traditional OS, we will frequently refer back to this section to explain what we are changing.

\subsection{Uncertainty Sampling}
\label{sec:unc_samp}

As constructed, the traditional OS requires estimates of every pulsar's noise auto-covariance, $\bm{P}_{a}$. For these, all deterministic signals, intrinsic pulsar noise, and common pulsar noise for each pulsar must be estimated in some way \citep{Anholm2009, Chamberlin2015, Vigeland2018}. These pulsar noise estimates often come from a quick PTA Bayesian analysis. Specifically, we get pulsar noise and common process estimates from a PTA common uncorrelated red-noise (CURN) search, which samples from a posterior with a similar likelihood to the one detailed in \autoref{eq:likelihood}, except that we set the cross-correlated terms $\bm{S}_{ab}$ of the covariance matrix in \autoref{eq:SigMatrix} to zero. We opt for a CURN search as it is roughly two orders-of-magnitude faster to analyze than a fully inter-pulsar correlated search, and provides reasonably accurate estimates for pulsar noise parameters in the presence of such a GWB-like process. 

Naively, we could use the maximum likelihood noise and CURN parameters from this MCMC analysis. This however does not account for uncertainty in these parameters as represented by the posterior spread, and may result in poor performance of the estimator. \citet{Vigeland2018} present a more robust methodology dubbed the Noise Marginalized Optimal Statistic (NMOS), in which they marginalize over the spread in parameter posteriors by computing the traditional OS over many draws from the MCMC chain. This results in a distribution of both $\hat{A}^2_{\mathrm{gw}}$ and $\sigma_{\hat{A}^2,0}$ 

The distribution of OS quantities given by the NMOS can be expressed as a Monte Carlo integral over the CURN posterior samples,
\begin{equation}
\begin{aligned}
    p(X | d) & = \int \delta(X-X_\eta) p(\eta | d) \, d\eta 
    \\
    & \approx \frac{1}{N} \sum_{i}^{N} \delta(X-X_{\eta_i}) 
    ,
\end{aligned}
\end{equation}
where $d$ represents the PTA data, $\eta$ is a vector of noise parameters for the PTA, $p(\eta|d)$ is the joint posterior distribution of these PTA noise parameters from the CURN search, $\{\eta_i\}_N$ are a set of $N$ random draws of these PTA noise parameters from $p(\eta|d)$, and $X\in\{\hat{A}^2_{\mathrm{gw}},\sigma_{\hat{A}^2,0}\}$ is either the estimated $\hat{A}^2_{\mathrm{gw}}$ or its uncertainty $\sigma_{\hat{A}^2,0}$ from the OS calculation, and $X_\eta$ is notation to indicate when these quantities have been computed for a specific choice of $\eta$. 

\begin{figure}
    \centering
    \includegraphics[width=\columnwidth]{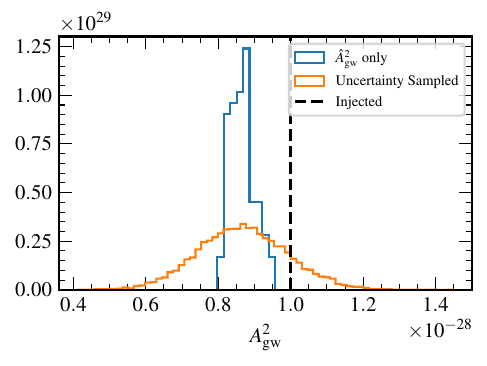}
    \caption{A normalized histogram of both the NMOS $\hat{A}^2_{\mathrm{gw}}$ estimator and the uncertainty-sampled estimator distributions for a realistic simulation with an injected GWB $A_{\mathrm{gw}} = 10^{-14}$. Ignoring the contributions of the NMOS's uncertainty results in a peaked distribution which excludes the injected GWB amplitude.}
    \label{fig:uncertainty_sampling}
\end{figure}

While the NMOS solves some problems with the traditional OS, it also introduces a new issue. Each vector of parameter estimates, $\eta$, creates its own $\hat{A}^2_{\mathrm{gw}}$ with Gaussian standard-deviation uncertainty $\sigma_{\hat{A}^2,0}$. In previous analyses the noise-marginalized distribution of $\hat{A}^2_{\mathrm{gw}}$ is shown while neglecting the $\sigma_{\hat{A}^2,0}$ values \citep{Arzoumanian2020,Vigeland2018}. More recent publications have omitted OS parameter estimation entirely \citep{NG15_evidence}, instead focusing on it as a detection statistic in which $\sigma_{\hat{A}^2,0}$ is only used to compute the SNR value for a particular $\eta$. As shown in \autoref{fig:uncertainty_sampling}, plotting only the distribution of $\hat{A}^2_{\mathrm{gw}}$ values and ignoring the underlying uncertainty in the estimator can cause narrow peaks in the estimated distribution, making the NMOS amplitude estimate seem more constrained than it really is.

To combat this problem, we propose an additional step in all OS pipelines, which we dub \textit{uncertainty sampling}. This step replaces each $\hat{A}^2_{\mathrm{gw}}$ estimate with many random draws from a Gaussian with mean, $\hat{A}^2_{\mathrm{gw}}$, and variance, $\sigma^2_{\hat{A}^2,0}$, implied by a given vector of pulsar parameter estimates, $\eta$.
\footnote{This is akin to using a Gaussian kernel density estimator where the width of each $\hat{A}^2$ point from NMOS sampling is determined by the point's corresponding uncertainty, $\sigma_{\hat{A}^2,0}$.} 
Importantly, while a Gaussian distribution is an appropriate approximation in usual circumstances, the true distribution is a generalized-$\chi^2$ \citep{Hazboun2023}.
We can frame this as an additional Monte-Carlo integral to the one already performed during the NMOS calculation, such that we can now approximate the full distribution on the GWB amplitude from the OS as
\begin{equation}
\begin{aligned}
    p(A^2_\mathrm{gw} | d) & = 
    \int p(A^2_\mathrm{gw} | \{\hat{A}^2_{\mathrm{gw}}, \sigma_{\hat{A}^2,0}\}_\eta) \, p(\eta | d) \, d\eta 
    \\
    & \approx \frac{1}{N} \sum_i^N p(A^2_\mathrm{gw} | \{\hat{A}^2_{\mathrm{gw}}, \sigma_{\hat{A}^2,0}\}_{\eta_i}) 
    \\
    & \approx \frac{1}{N\times M} \sum_i^N \sum_j^M \delta(A^2_\mathrm{gw} - Y_{\eta_i,j}),
\end{aligned}
\end{equation}
where $\{Y_{\eta_i,j}\}_M \sim N(\hat{A}^2_{\mathrm{gw}}, \sigma_{\hat{A}^2,0})_{\eta_i}$ are $M$ samples from the uncertainty distribution of the OS (assumed to be Gaussian) with mean and standard deviation given by $\hat{A}^2_{\mathrm{gw}}$ and $\sigma_{\hat{A}^2,0}$, which depends on a given set of PTA noise parameters $\eta_i$.

Using our realistic simulations, we find that $10^3$ NMOS samples, and using uncertainty sampling of 100 draws per NMOS sample, was sufficient in stabilizing the final estimator distribution. However, the exact numbers for these vary depending on the data set, and the stability of the distribution should always be checked. In this case, we did so by comparing the full uncertainty sampled $A^2_{\mathrm{gw}}$ distribution with one made with a smaller portion of those of the same distribution, e.g., 90\% of the $A^2_{\mathrm{gw}}$ distribution. If these two distributions closely match, the solution is likely stable. 

With uncertainty sampling, we now have a distribution of possible $A^2_{\mathrm{gw}}$ values that accounts for both the spread from the PTA noise parameter estimation and the spread in the GWB amplitude estimation. Shown in orange in \autoref{fig:uncertainty_sampling}, we see that the widening of the distribution now covers the injected value of the simulation. All subsequent analysis in this paper will be performed using uncertainty sampling, unless otherwise noted.

\subsection{Parameter estimation}
\label{sec:par_est}

As so far discussed, the traditional OS is designed to work for PTA data in the weak-signal regime, defined such that the contribution of the GWB to the total noise is negligible in all measured frequencies \citep{Siemens2013}. This assumption leads to using a null-hypothesis diagonal pulsar-pair covariance matrix $\bm{C}_{0}$ in \autoref{eq:Ahat}. When evaluating the traditional OS in the weak-signal regime, the amplitude estimator is unbiased \citep{Vigeland2018}. 

The problem, however, is that some PTA datasets are no longer in the weak-signal regime, as the GWB signal is becoming the dominant contribution to the total red noise in the lowest frequencies of many pulsars \citep{NG15_evidence}. This was first proposed in the LIGO case in \citet{Allen1999}, and was revisited for the pulsar timing case in \citet{allen-romano2023}. The first evidence of this issue in realistic simulations was shown as a part of the NANOGrav 15 year publications \citep{Johnson2023} and as such, was included within the analysis in Figure 1 of the NANOGrav 15 year evidence paper \citep{NG15_evidence}. The way to solve this problem is to recalculate our covariance matrix without any simplifying assumptions,
\begin{equation}
    \bm{C} = \bm{C}_{ab,cd} = \expect{\rho_{ab} \, \rho_{cd}} - \expect{\rho_{ab}} \expect{\rho_{cd}} 
    ,
\end{equation}
in larger signal regimes, the correlated GWB signals in the $\bm{S}_{ab}$ terms we originally assumed to be zero in Appendix \ref{sec:os_sigma_calc} become more significant. Ignoring these GWB contributions results in an estimator with a variance that is too small \citep{allen-romano2023}. This was also influential in the recent NANOGrav 15-year GWB analysis, in which it was found that neglecting these terms results in $20-40\%$ smaller uncertainties \citep{NG15_evidence}. The solution to this is a new technique of including the pulsar-pair covariance \citep{Romano2021, allen-romano2023, Johnson2023}, which involves a replacement of the approximate diagonal $\bm{C}_{0}$ with a its exact dense form $\bm{C}$ in \autoref{eq:Ahat} and \autoref{eq:sigma_ahat}. While mathematically straightforward, calculating all elements of this covariance matrix can be computationally expensive without simplifying strategies. We use the covariance matrices described in \citet{Johnson2023} and apply the same rank-reducing strategies found in \citet{pol2022anisotropy} to drastically reduce the computational resource needs. We detail the calculation of the pulsar-pair covariance matrix elements in Appendix \ref{sec:pair_covariance_calc}.

A key step in calculating the pair covariance matrix is the inclusion of an assumed GWB amplitude (see equations \ref{eq:pc_nomatch}, \ref{eq:pc_onematch}, and \ref{eq:pc_twomatch}). This presents a somewhat circular problem, as we must assume a GWB amplitude to then measure the GWB amplitude. \citet{allen-romano2023} proposes a treatment with the assumed amplitude and estimated amplitude as separate quantities, and use $\chi^2$-fitting to find regions in which the assumed and measured amplitudes agree.  Attempting this solution however, we find that the inversion of the pulsar-pair covariance matrix can become numerically unstable in some cases, making it impossible to find regions in which the assumed and measured amplitudes agreed. The scenarios and regimes when this occurs have not yet been fully explored and are needed to implement this proposed solution.

We instead decide to sidestep this issue by implementing the same strategies that the traditional OS does. Rather than varying the assumed GWB amplitude, we set this value to the value found in the preliminary Bayesian CURN search. This scheme is consistent with the way the current implementation assumes the values of the auto-correlations in the covariance matrix. In our testing with simulated data, this scheme resulted in estimated GWB amplitudes that are consistent with the injected amplitudes, even in strong signal simulations.

We compare the performance of the traditional OS with and without pulsar-pair covariance using Percentile-Percentile (P-P) plots. We created 100 realistic simulated datasets with a GWB power-law amplitude of $A^2_{\mathrm{gw}} = 10^{-14}$ (additional details about these simulations are given in \autoref{sec:sim_design}). This amplitude is intentionally chosen to be far stronger than is seen in current datasets (NANOGrav has a median amplitude of $2.4 \times 10^{-15}$ \citep{NG15_evidence}). This ensures that the GWB dominates in all 10 frequencies used in the analysis in order to specifically strain the abilities of the traditional OS with and without pulsar-pair covariance. 

\begin{figure}
    \centering
    \includegraphics[width=\columnwidth]{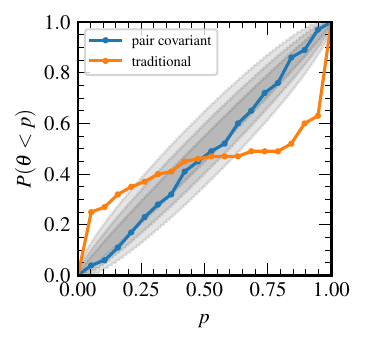}
    \caption{P-P plot comparing the traditional OS with and without pulsar-pair covariance incorporated. The pair covariance calculation assumes injected parameter values. Each line represents the same 100 realistic simulations with an injected GWB amplitude of $10^{-14}$. The shaded regions indicate the 1$\sigma$, 2$\sigma$, and 3$\sigma$ confidence contours for an unbiased estimator.}  
    \label{fig:12p5_os_v_pcos_pp}
\end{figure}

For each simulation we used the NMOS by directly providing it with the injected parameter values for the simulation, effectively assuming that our CURN analysis parameter recovery is perfect. We then used uncertainty sampling to create a distribution of $A^2_{\mathrm{gw}}$ values, after which we found the percentile at which the injected amplitude lies. In \autoref{fig:12p5_os_v_pcos_pp}, the P-P plot shows the cumulative distribution function of the injected-amplitude percentiles from each simulation for the traditional OS with and without pulsar-pair covariance modeling.
We clearly see the impact of incorporating pulsar-pair covariance, as this new estimator remains within the $95\%$ confidence region throughout, while without it the estimator shows a sideways `S' shape, which is characteristic of an underestimated uncertainty \citep{Lamb2023}.

\subsection{Detection statistic}

The second major use case for the optimal statistic is to gauge the significance of inter-pulsar correlations induced by a GWB signal. We can compute a signal-to-noise ratio (SNR) by taking the ratio of our estimator $\hat A_{\mathrm{gw}}^2$ and its uncertainty calculated assuming the null hypothesis:
\begin{equation}
    \mathrm{SNR} = 
    \frac{\hat{A}^2_{\mathrm{gw}}}{\sigma_{\hat{A}^2,0}}. 
\end{equation}
The traditional OS SNR is often used as an informal proxy for the significance of cross-correlations in terms of standard deviations away from zero; however the null hypothesis distribution of the traditional OS is formally non-Gaussian, and therefore the significance must be calibrated empirically or through specification of the true generalized-$\chi^2$ null hypothesis distribution \citep{Hazboun2023}.

Expanding this scheme with the NMOS is trivial, as each vector of PTA parameter estimates results in an SNR. Over many draws of PTA parameter estimates, we can construct a distribution of SNRs where the median more closely follows the injected-value SNR \citep{Vigeland2018}. Unlike in the case of parameter estimation, the SNR calculation works well without the inclusion of pulsar-pair covariance, since it will be calibrated anyway through data augmentation and bootstrapping techniques. The two current methods of calibration are numerical estimation through schemes like sky scrambles and phase shifts \citep{Cornish2016, Taylor2017}, or analytic derivation by calculating the generalized-$\chi^2$ distribution that describes the true shape of the null distribution \citep{Hazboun2023}.

\section{A New Per-Frequency Optimal Statistic}
\label{sec:PFOS_derivation}

The primary goal of this work is to further generalize the traditional OS to work for any GWB spectral shape, not just power-law behavior. We call the method presented here the Per-Frequency Optimal Statistic (PFOS). The traditional OS was constructed to measure the correlated amplitude and SNR of a power-law spectrum GWB signal, using a unit-amplitude power-law spectral template for the GWB, $\hat{\phi} = \phi / A^2_{\mathrm{gw}}$, where $\phi$ represents the power-law PSD of the GWB in \autoref{eq:powerlaw}. This broadband estimation condenses the correlated information from each frequency into a single, more sensitive measurement \citep{Anholm2009}. Recalling the rank-reduced forms in \autoref{eq:rho_rank_reduced} and \autoref{eq:sigma_rank_reduced} for the cross-correlation estimator, we can change the spectrum which that estimator assumes by simply swapping $\hat{\phi}$ with another spectral template. 

Here we change the spectral template to be entirely agnostic to the spectral shape. Since the traditional OS already deals in discrete Fourier frequencies in both the $\bm{X}_a$ and $\bm{Z}_a$ terms, we replace the unit-amplitude Fourier domain covariance matrix, $\hat{\phi}$, with a frequency-selector matrix, $\Tilde{\phi}(f_k)$, which multiplies and adds the adjacent sine and cosine components of the two pulsars at the specified frequency. This frequency-selector matrix will retain the same shape as $\phi$, ($2k_\mathrm{max} \times 2k_\mathrm{max}$) elements and still be diagonal, however, all of the elements of this diagonal matrix will be zero except for those which correspond to frequency $f_k$. This is best shown through examples,
\begin{equation}
\begin{aligned} 
    \Tilde{\phi}(f_1) & = \mathrm{diag}( 1, 1, 0, 0, ... , 0, 0) , \\ 
    \Tilde{\phi}(f_2) & = \mathrm{diag}( 0, 0, 1, 1, ... , 0, 0) , \\ 
    \Tilde{\phi}(f_{k_\mathrm{max}}) & = \mathrm{diag}( 0, 0, 0, 0, ... , 1, 1) , \\ 
\end{aligned}
\end{equation}
where $f_k$ represents the frequency to be analyzed (which must be a multiple of the lowest PTA frequency). However, swapping the spectral template from $\hat{\phi}$ to $\Tilde{\phi}(f_k)$ does introduce a new assumption that PTA GW frequency
measurements are independent of one another; we test this assumption in \autoref{subsec:broadband_recovery}. 

We now construct a new pair-wise cross-correlation estimator at the specified frequency,
\begin{equation}
    \rho_{ab}(f_k) = \mathcal{N}_{ab}(f_k) \, \bm{X}^T_a \Tilde{\phi}(f_k) \bm{X}_b
    ,
    \label{eq:pfos_pairwise_unnomed}
\end{equation}
where $\mathcal{N}_{ab}(f_k)$ is now a frequency-dependent normalization, defined such that that the expectation is the ORF-independent GWB PSD, $S(f_k)$, at frequency $f_k$, $\expect{\rho_{ab}(f_k)} = \Gamma_{ab} \, S(f_k)$. This calculation is presented in full in Appendix \ref{sec:PFOS_normalization}. The resulting estimator is then
\begin{equation}
    \rho_{ab}(f_k) = 
    \frac{ \bm{X}^T_a \Tilde{\phi}(f_k) \bm{X}_b }
    {\trace{ \bm{Z}_a \Tilde{\phi}(f_k) \bm{Z}_b \Phi(f_k)}} 
    ,
    \label{eq:rhof}
\end{equation}
and the variance in this estimator---calculated in Appendix \ref{sec:pfos_uncertainty}---is
\begin{equation}
    \sigma_{ab,0}(f_k)^2 = 
    \frac{ \trace{\bm{Z}_a \Tilde{\phi}(f_k) \bm{Z}_b \Tilde{\phi}(f_k)} }
    {\trace{ \bm{Z}_a \Tilde{\phi}(f_k) \bm{Z}_b \Phi(f_k)}^2} 
    .
    \label{eq:sigmaf}
\end{equation}

In both of these equations, we introduce a new matrix quantity $\Phi(f_k) = \phi / S(f_k)$. This represents the shape of the Fourier-domain covariance matrix with unit-value at frequency $f_k$. This template will be unique for each frequency measured and allows us to measure each frequency one at a time. Similar to the traditional OS, the uncertainties in \autoref{eq:sigmaf} are calculated under the null hypothesis where $\expect{\rho_{ab}(f_k)} = 0$. Since the normalization term introduces some broadband dependence to the PFOS, we refer to this as the \textit{broadband-normalized PFOS} (or just the PFOS).

Including $\Phi(f_k)$ may seem counterintuitive, as it requires information on the spectrum we are currently trying to estimate. However, this is similar to the problem of incorporating $A^2_{\mathrm{gw}}$ as an assumed amplitude when using pulsar-pair covariance with the traditional OS. We employ the same strategy that was discussed in \autoref{sec:par_est} where we use the CURN estimates of the GWB as the assumed spectral shape. As we will soon show, this scheme produces estimates that are consistent with injection values in our simulations while also allowing us to fit frequencies one at a time and keeping our covariance matrices smaller and more stable. Details about effects resulting from the choice of a CURN model can be found in \autoref{subsec:per_frequency_recovery} and \autoref{sec:discussion}. 

We can now estimate the GWB PSD, $S(f_k)$, using pair-wise correlation information from the full PTA, such that
\begin{equation}
    \hat{S}(f_k) = 
    \left[ \vec{\Gamma}^T \bm{C}(f_k)^{-1} \vec{\Gamma} \right]^{-1}
    \vec{\Gamma}^T \bm{C}(f_k)^{-1} \vec{\rho(f_k)}
    ,
    \label{eq:S_f}
\end{equation}
and the corresponding uncertainty at each frequency can be written as,
\begin{equation}
    \sigma_{\hat{S}(f_k),{0}} =
    \left[ \vec{\Gamma}^T \bm{C}(f_k)^{-1} \vec{\Gamma} \right]^{-1/2}
    .
    \label{eq:sigma_sf}
\end{equation}

In these equations, we again express our pair-wise estimators as a vector of all pairs for a particular frequency $f_k$ with a corresponding vector of ORF values. The pulsar-pair covariance matrix for a particular frequency is then written as,
\begin{equation}
    \bm{C}(f_k) = 
    \expect{\rho_{ab}(f_k) \rho_{cd}(f_k)} - 
    \expect{\rho_{ab}(f_k)}\expect{\rho_{cd}(f_k)}.
\end{equation}

Similar to the traditional OS, we can choose to incorporate pulsar-pair covariance or not. If we do not then the covariance matrix is again diagonal, leading to $\bm{C}_{ab,ab}(f_k) = \delta_{ac} \delta_{bd} \, \sigma_{ab,0}(f_k)^2$. Including pair covariance requires calculating all of the elements from the $\bm{C}_{ab,cd}(f_k)$ matrix, which can be found in Appendix \ref{sec:pair_covariance_calc}.

\subsection{Frequency-independent simplification}
\label{subsec:narrowband}

We can make an additional simplification to the PFOS method if we assume that all of our pulsars measure each GW frequency independently of each other frequency. This assumption is valid if every pulsar is timed with equal cadences and contains no data gaps. Under this additional assumption, the $Z$ matrices defined in equation \autoref{eq:Z} become frequency block-diagonal. With a frequency block-diagonal matrix, the normalization calculated in \autoref{sec:PFOS_normalization} become independent of the spectral shape matrix $\Phi(f_k)$. We can then replace this spectral shape matrix with a much simpler frequency-selector matrix $\Tilde{\phi}(f_k)$. As this also re-normalizes the PFOS as a frequency-independent narrowband estimator, we will refer to this form as the \textit{narrowband-normalized PFOS}.

Substituting this frequency-selector matrix into \autoref{eq:rhof} and \autoref{eq:sigmaf} leads us to write our cross-correlation estimators and associated variances as
\begin{equation}
    \rho_{ab,\mathrm{narrow}}(f_k) = 
    \frac{ \bm{X}^T_a \Tilde{\phi}(f_k) \bm{X}_b }
    {\trace{ \bm{Z}_a \Tilde{\phi}(f_k) \bm{Z}_b \Tilde{\phi}(f_k)}} 
    ,   
    \label{eq:rho_narrow}
\end{equation}
\begin{equation}
    \sigma_{ab,0,\mathrm{narrow}}(f_k)^2 = 
    \trace{ \bm{Z}_a \Tilde{\phi}(f_k) \bm{Z}_b \Tilde{\phi}(f_k)}^{-1}
    .    
    \label{eq:sigma_narrow}
\end{equation}
We can then still use \autoref{eq:S_f} and \autoref{eq:sigma_sf} to estimate $S(f_k)$. 

This narrowband-normalized PFOS looks remarkably like the traditional OS's estimators and variances in \autoref{eq:rho_rank_reduced} and \autoref{eq:sigma_rank_reduced}. This simplicity leads us to initially develop this form before a more rigorous calculation of the normalization was completed.

While the narrowband-normalized PFOS benefits from the lack of an estimated spectral shape, this simplicity comes at a cost. When making this assumption in data sets where these GW frequencies are not independent, the narrowband normalization ignores contributions from other frequencies, leading to overestimation of the GWB spectrum. These problems become more pronounced in larger signal scenarios.

Based on our testing, this simplification can work well in more ideal data sets where individual pulsars have similar observing cadence and length. However, as we will show in \autoref{subsec:per_frequency_recovery}, using more realistic data shows worse performance than the broadband-normalized PFOS. 
Additionally, even in data sets where the narrowband-normalized PFOS simplification can be applied, the broadband-normalized PFOS will result in identical answers. For these reasons, while we elect to show this as possible simplification, we \textbf{strongly} recommend using the broadband-normalized PFOS instead.

\subsection{Signal-to-Noise ratio interpretations}

Finally, it is also possible to use the different flavors of PFOS to calculate SNRs at each frequency. Similar to the traditional OS, the PFOS's null hypothesis is the absence inter-pulsar correlations. However, the null hypothesis should be interpreted differently depending on whether the broadband-normalized or narrowband-normalized PFOS is used. 

The broadband-normalized PFOS implies a null hypothesis in which one assumes that no measurable correlations exist in a particular frequency bin, but permits them to be present in other bins. This is less biased in cases where there are significant correlations in other frequency bins, making it possible to isolate the significance of a single frequency bin in a detectable spectrum. 

By contrast, the narrowband-normalized PFOS implies a null hypothesis in which one instead assumes that no frequency bins whatsoever have measurable correlations. Hence the narrowband-normalized PFOS null hypothesis is most useful in cases where there are not significant correlations present at any frequency, and as such we are trying to estimate the significance in a single frequency bin. These interpretations and the potential for other interpretations of the null hypothesis should be deliberated upon on a case-by-case basis, and should always be calibrated through empirical techniques \citep{Taylor2017, Cornish2016}.

\section{Simulation Design}
\label{sec:sim_design}

We design sets of simulated datasets in order to test the efficacy of the PFOS. These simulated data sets are all created using the \texttt{toasim} methods found within the \texttt{libstempo} python package \citep{libstempo}. Each simulation is based on the pulsars of NANOGrav's 12.5 year data set \citep{Arzoumanian2020}, including their observational schedules and noise characteristics. We use many of the same simulation methods found in \citet{Pol2021astroMilestones}. Briefly, we use all $45$ pulsars which have timing baselines longer than 3 years, where the full full PTA baseline is $T_{\mathrm{span}} = 12.9$~years. We then condense the many near-simultaneous TOA measurements resulting from narrow-band timing into single epoch-averaged TOAs. This reduces the original $410,064$ TOA measurements into just $6,244$ with a median TOA uncertainty shrinking from $2.0 \times 10^{-6}\,\mathrm{s}$ to $2.9 \times 10^{-7}\,\mathrm{s}$. For each simulation, we opted to search in the 10 lowest Fourier bins of our PTA, corresponding to $f_k = n/T_{\mathrm{span}}$ where $n\in[1,10]$. These changes simplify our simulated datasets to make injections and analyses faster, while still retaining realistic data quality features such as multi-year gaps in TOAs for some pulsars, observations from multiple telescopes, and differences in pulsar timing baselines. 

For each pulsar, we simulate timing data that includes white noise, intrinsic red noise, and a GWB signal that is correlated among pulsars. The white noise in each pulsar is drawn according to the TOA uncertainties. The intrinsic red noise is generated as a time-domain realization  of a process with a power-law power spectrum, with power-law parameters matching the maximum likelihood values found in \citet{Arzoumanian2020}. For our analyses, we opt to fix these intrinsic red noise power-law parameters in our analyses. This eliminates the need to simultaneously search for these parameters, vastly improving the speed and sampling performance for the initial Bayesian CURN search upon which the OS and its variants rely.  We verified in a small number of simulation studies that the behavior when we fixed versus varied the intrinsic red-noise parameters was similar. The major difference was the larger uncertainty in the final PFOS-derived quantities compared to when intrinsic red noise was held fixed.

Finally, to each simulated dataset we add a GWB signal that is correlated across all pulsars according to the Hellings \& Downs curve, with a power-law power spectrum that can be parameterized by the amplitude and spectral index found in \autoref{eq:powerlaw}. All of our simulations have a spectral index of $\gamma = 13/3$.  We split our $300$ total simulations into three groups of $100$, where each group has a different injected power-law GWB amplitude:  $A_{\mathrm{gw}}\in[10^{-16}, 10^{-15}, 10^{-14}]$. These correspond to a median NMOS SNR of $0.02$, $1.91$, and $7.43$, respectively, creating the conditions of the weak, intermediate, and strong signal regimes. These regimes  are described by \citet{Siemens2013} in terms of the ratio of the GWB PSD to those of other noise sources. In the weak signal regime, all GW frequencies analyzed are noise dominated; the intermediate regime has some fraction of frequencies dominated by the GWB; and the strong regime has all analyzed frequencies dominated by the GWB.

We perform initial Bayesian analyses of each simulation assuming a power-law CURN approximate model for the GWB signal (i.e., no inter-pulsar correlations modeled) with two parameters corresponding to power-law spectral index, $\gamma$, and the amplitude at $f_{\mathrm{yr}}$. We find that this model produces PTA parameter estimates that lead to better and more consistent performance when passed to the PFOS framework than when we use a more agnostic free-spectrum CURN search, as variations in spectral shape between samples is far more consistent. See \autoref{sec:discussion} for further discussion on the choice of model for the initial pilot Bayesian characterization of the PTA noise.

\section{Results}
\label{sec:method_test}

In this section we demonstrate the accuracy of the broadband-normalized PFOS---henceforth referred to as the PFOS---in our different simulated dataset scenarios. This section is split into four parts. First, we analyze the capabilities of the PFOS on a single case-study data set, both for measuring the PSD of the GWB and for reconstructing the spatial correlations at each frequency. Second, using the suite of simulations discussed in \autoref{sec:sim_design}, we assess the accuracy of PFOS-derived PSD measurements at each frequency. Third, we test broadband power-law spectral fitting with the PFOS products in order to compare the performance to that of the traditional OS, in which a broadband power-law model is assumed from the beginning. Finally, we show results from a simulation in which an isotropic stochastic GWB signal as well as a prominent supermassive black-hole binary GW signal have been injected, demonstrating the prospects for excess-power--based searches for individual sources with the PFOS.

\subsection{Case study on a single data set}
\label{subsec:data_products}

\begin{figure}
    \centering
    \includegraphics[width=\columnwidth]{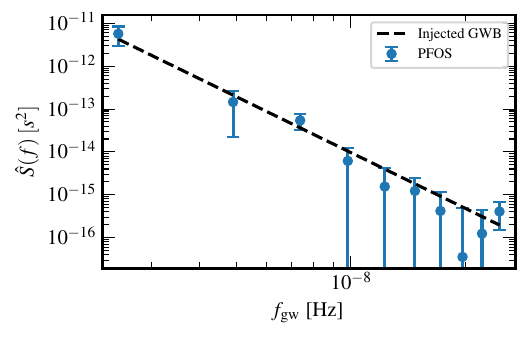}
    \caption{The recovered pulsar-pair correlated GWB spectrum from the (broadband-normalized) PFOS given the injected simulation parameters with point estimates on $S(f_k)$ and associated 1-sigma error bars. The black dashed line indicates the injected GWB PSD.}
    \label{fig:pfos_spectrum}
\end{figure}

For our first test we analyze a single simulation with an injected GWB amplitude of $A_{\mathrm{gw}}=10^{-14}$ at $f={\rm yr}^{-1}$, where the PFOS calculation is performed with the inclusion of pulsar-pair covariance, and in which we supply the injected noise and signal hyper-parameters. This gives the reconstructed spectrum of timing variance induced by the GWB shown in \autoref{fig:pfos_spectrum}. All but one frequency contains the injected GWB spectrum within their $1\sigma$, where the eighth frequency contains the injected value just outside this at the $1.03\sigma$ level.

As with the traditional OS, the PFOS allows one to construct a binned estimator to visualize the cross-correlated power as a function of pulsar angular separations. However, the PFOS's ability to separate correlations between frequencies allows us to create these binned estimators at each frequency, enabling more precise examinations of the structure of cross-correlations in our data. We use methods detailed in \citet{allen-romano2023} and \citet{Johnson2023} to create a binned estimator of cross-correlations with corresponding uncertainty,
\begin{equation}
\begin{aligned}
    \rho_{\mathrm{opt},i}(f_k) &=
    \Gamma_{\xi_i} \, \left( \vec{\Gamma_{i}}^T [\bm{C}_i(f_k)]^{-1} \vec{\Gamma_{i}} \right)^{-1} 
    \nonumber \\   
    &\hspace{0.5in}\times \vec{\Gamma_{i}}^T [\bm{C}_i(f_k)]^{-1} \vec{\rho_i(f_k)}
    ,
\end{aligned}
\end{equation}
\begin{equation}
    \sigma_{\mathrm{opt},i}(f_k) =
    \left( \vec{\Gamma_{i}}^T [\bm{C}_{i}(f_k)]^{-1} \vec{\Gamma}_{i} \right)^{-1/2}.
\end{equation}
In these equations, $i$ is an index selector which represents the subset of pulsar pairs for a particular angular separation bin, $\vec\Gamma_{i}$ is a vector of ORF values for each pulsar pair within bin $i$, $\vec\rho_{i}(f_k)$ is a vector of pulsar-pair correlated PSD estimators within bin $i$, $\bm{C}_{i}(f_k)$ is the pulsar-pair covariance matrix for frequency $f_k$ for pulsar-pairs within angular separation bin $i$, and $\Gamma_{\xi_i}$ is the ORF of the average angular separation of the pulsar pairs in bin $i$. 

\begin{figure}
    \centering
    \includegraphics[width=\columnwidth]{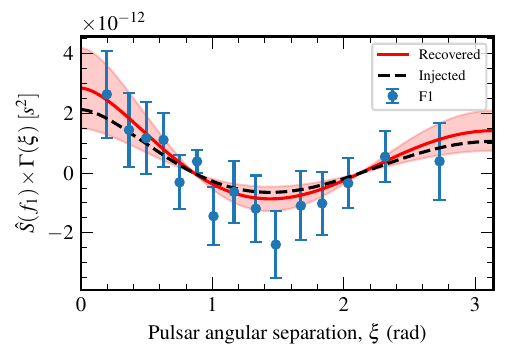}
    \caption{The recovered binned pulsar pair correlated power for the first GW frequency ($f_1=2.45$~nHz) for the (broadband-normalized) PFOS. The dashed black line represents the injected GWB signal. The solid red represents the best fit signal. The shaded red region represents the 1-sigma uncertainty on the best fit signal. The blue points represent the binned cross-correlation measurements, where the bins were chosen so that there are the same number of pulsar pairs in each bin.}
    \label{fig:pfos_f1}
\end{figure}

The resulting set of estimators for the first GW frequency ($f_1=2.45$~nHz) for our simulation using the PFOS is shown in \autoref{fig:pfos_f1}. Note that---despite being suppressed in this visualization---the bin-wise estimators are covariant. Additionally, the correlated PSD estimate and corresponding uncertainty represented in red are computed using the full covariance between each pulsar pair. It would be straightforward to incorporate the newly developed Multiple Component Optimal Statistic (MCOS) to search for many correlated processes simultaneously at specific frequencies \citep{Sardesai2023}.

\begin{figure}
    \centering
    \includegraphics[width=\columnwidth]{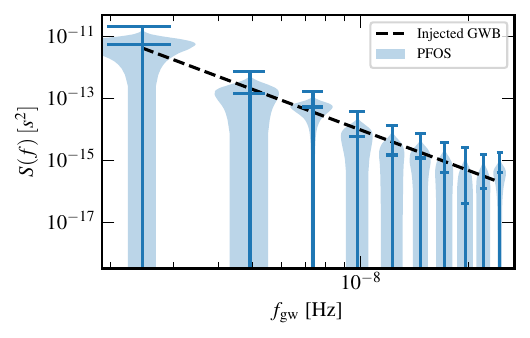}
    \caption{The recovered GWB spectrum from the (broadband-normalized) PFOS using CURN sampling for parameter estimation. The distributions on each frequency represent the uncertainty sampled noise marginalized distribution on the PFOS' recovery.}
    \label{fig:pfos_unc_samp}
\end{figure}

Once we incorporate noise marginalization and uncertainty sampling, we create a distribution of $\hat{S}(f_k)$ that will more accurately represent the total uncertainty in the estimator. \autoref{fig:pfos_unc_samp} shows this in a violin plot with the medians represented as horizontal lines and extrema represented by caps. Since we are looking at a full distribution of estimators rather than $1\sigma$ error bars, it is typical to see each violin include both the injected value and zero. While negative power is non-physical, the nature of linear fitting with \autoref{eq:S_f} allows for any real-valued ORF amplitude. In this case, the negative values represent an inverted HD curve, which, if small, we interpret as noise fluctuations.

\subsection{Per-frequency PSD recovery}
\label{subsec:per_frequency_recovery}

We now more systematically test the PSD recovery frequency-by-frequency using the suite of simulations described in Section \autoref{sec:sim_design}. Each simulation's CURN posterior distribution is sampled sufficiently with MCMC techniques before $10^3$ parameter vectors are randomly selected to pass through the noise-marginalized PFOS. From the distributions of $\hat{S}(f_k)$ and $\sigma_{\hat{S}(f_k)}$, we use uncertainty sampling with $10^2$ draws from each Gaussian to approximate the full distribution of $S(f_k)$, thereby incorporating both the Bayesian CURN parameter uncertainty and the PFOS estimator uncertainty. We then find the percentiles at which the injected GWB amplitude lies within these full distributions, and use these to construct P-P plots for each frequency and each tested GWB amplitude. The behavior of these P-P plots can be characterized with respect to a diagonal, unbiased estimator \citep[see, e.g.,][and references therein]{Lamb2023}.

\begin{figure}
    \centering
    \includegraphics[width=\columnwidth]{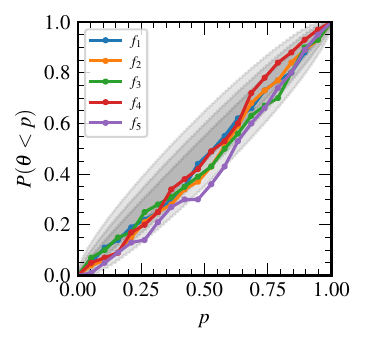}
    \caption{Individual frequency P-P plots at a power-law amplitude of $A_{\mathrm{gw}}=10^{-14}$ for the (broadband-normalized) PFOS using pair-covariant, uncertainty sampling, and noise marginalization from a Bayesian CURN search. The shaded regions indicate the 1, 2, and 3 sigma contours from a perfectly unbiased estimator for these 100 simulations. Each frequency is denoted by its multiple of $T_{\rm span}^{-1}$.}
    \label{fig:PFOS_12p5_VG_F_1e-14}
\end{figure}

Shown in \autoref{fig:PFOS_12p5_VG_F_1e-14} are P-P plots representing the first five frequencies of the $100$ simulations with the largest GWB amplitude of $A_{\mathrm{gw}}=10^{-14}$. These strong-signal regime simulations show the largest bias compared with the weak and intermediate signal regimes, and are representative of a worst-case scenario for the PFOS. 
This figure shows a general trend of underestimation for all frequencies and percentiles. This is an indication of an underestimated mean \citep{Lamb2023}. We also see that while the first four frequencies remain within the $3\sigma$ contours, the fifth frequency briefly departs this contour at the $50^{\rm th}$ percentile.
Higher frequencies (not shown) exhibit the same slightly underestimated but relatively unbiased behavior as the lowest four frequencies. The increase in bias at the fifth frequency around the $50^{\rm th}$ percentile is also present with weaker GWB amplitudes but are less pronounced. 
Notably, in more ideal simulations the PFOS recovery has no such problems with any frequencies. This isolation of a single problematic frequency in realistic data may indicate an issue with the PFOS recovery with some PTA configurations. Differences between these and more ideal simulations including irregular timing cadences, timing gaps, and non-uniform timing baselines, may cause a break down in the frequency independence of the PFOS. More work must be done to determine exactly when these discrepancies are problematic. However, as \autoref{subsec:broadband_recovery} will soon show, the effects of this bias are small enough to create consistent measurements with the traditional pipeline.
It is expected that the PFOS will be at least somewhat biased, as to be completely unbiased would require knowledge of the exact shape of the GWB spectrum beforehand.

\begin{figure}
    \centering
    \includegraphics[width=\columnwidth]{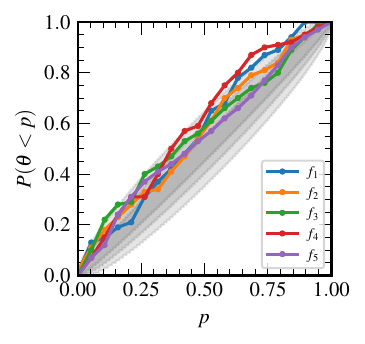}
    \caption{
    Individual frequency P-P plots for a power-law amplitude of $A_{\mathrm{gw}}=10^{-14}$ for the narrowband-normalized PFOS using pair-covariance, uncertainty sampling, and noise marginalization from a Bayesian CURN search. The shaded regions indicate the $1$, $2$, and $3\sigma$ contours from a perfectly unbiased estimator for these $100$ simulations. Each frequency is denoted by its multiple of $T_\mathrm{span}^{-1}$.
    }    
    \label{fig:PFOS_rad_12p5_VG_F_1e-14}
\end{figure}

We also test the narrowband-normalized PFOS which is agnostic to the assumed spectral shape. 
Since the narrowband-normalized PFOS as constructed is mathematically biased in non-ideal data, we expect this method to perform worse overall. \autoref{fig:PFOS_rad_12p5_VG_F_1e-14} shows a P-P plot of the first 5 GW frequencies of the same simulations. This time analyzed using the narrowband-normalized PFOS.
Surprisingly, each frequency remains broadly consistent within the $3\sigma$ contours with only a minor deviation in the fourth frequency. 
We note however that the mean at every frequency is systematically overestimated. This is expected behavior as the smaller overall weighting of the normalization shifts these estimates to higher values.
For these reasons, the more accurate and robust solutions provided by the broadband-normalized PFOS are preferred.

\subsection{Broadband spectrum recovery}
\label{subsec:broadband_recovery}

In addition to agnostic spectral recovery, we also test the performance of fitting the PFOS' PSD estimates with a parameterized spectral model. We opt for a power-law so that we can directly compare the PFOS model fitting to the traditional OS. As with the individual frequency tests, we source our CURN power-law estimates from Bayesian CURN analyses. All three methods incorporate pulsar-pair covariance and, to provide a fair comparison, are provided with the same vectors of parameter estimates. Given all 10 frequencies of the individual PSD estimates resulting from the two different PFOS methods, we carry out a least-squares fit for a power-law amplitude (referenced to a frequency of $f={\rm yr}^{-1}$) with the spectral index $\gamma$ set to the value found in the Bayesian CURN power-law spectral index parameter. This exactly mimics the process of the traditional OS's fitting. With noise marginalization and uncertainty sampling for both PFOS methods and traditional OS, we find the percentiles at which the injected GWB amplitude lies in each of our $100$ simulations.

\begin{figure}
    \centering
    \includegraphics[width=\columnwidth]{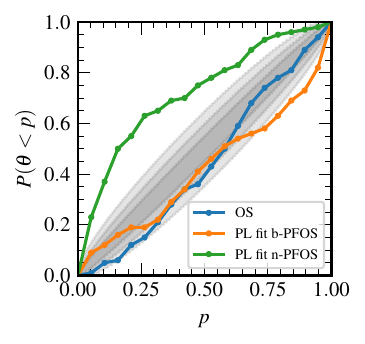}
    \caption{Broadband P-P plots for a power-law amplitude of $A_{\mathrm{gw}}=10^{-14}$ for a linear power-law fit on the pair-covariant broadband-normalized PFOS (b-PFOS), narrowband-normalized PFOS (n-PFOS), and pair-covariant OS using uncertainty sampling and noise marginalization from a Bayesian CURN search. The shaded regions indicate the 1, 2, and 3 sigma contours from a perfectly unbiased estimator for these 100 simulations.}
    \label{fig:PFOS_v_OS}
\end{figure}

The results for $A_{\mathrm{gw}}=10^{-14}$ are shown in \autoref{fig:PFOS_v_OS}. As with the individual frequency analysis, the largest GWB amplitude produces the greatest deviations from an unbiased estimator, such that \autoref{fig:PFOS_v_OS} again represents a worst-case scenario. We see that the traditional OS with pulsar-pair covariance, labeled as ``OS'', performs well and remains within the $3\sigma$ confidence contours at all percentiles, even in this strong signal regime, though it does have some minor bias at lower percentiles. By contrast, the power-law fit on the broadband-normalized PFOS analysis, labeled as ``PL fit b-PFOS'', contain other minor deviations. There is a slight sag, especially with higher percentiles but surprisingly does better than the traditional OS in lower percentiles. The fifth frequency problem and general underestimation shown in \autoref{fig:PFOS_12p5_VG_F_1e-14} may be causing some of the discrepancies in higher percentiles, as more idealized simulations with even cadences show even better agreement between the two methods. Regardless, these P-P plots still show remarkable agreement despite the high strength of the GWB used.

Unlike the previous two cases, the narrowband-normalized PFOS, labeled as ``PL fit n-PFOS'', vastly overestimates the injection at all percentiles while the uncertainty estimates were nearly identical to the broadband-normalized. This is a result of using the individual frequency overestimates which are then compounded into a worse broadband spectral fit. This further reinforces our recommendations to use the broadband-normalized PFOS instead.

These tests demonstrate that the PFOS's ability to accurately measure the GWB PSD is sufficient to then use in a full spectral model fit. 
Despite the fact that we remove both the power-law assumption of the traditional OS and any frequency covariance information, 
the broadband-normalized PFOS can produce nearly equivalent results when both methods correctly account for pulsar-pair covariance.

\subsection{Stochastic background plus single GW signal}
\label{subsec:GWB+CW}

As a final test, we explore the potential for the PFOS to provide early indicators of a single supermassive black-hole binary system's GW signal resounding above the stochastic GWB. For this test, we opt for a pair of PTA simulations: one with only a GWB signal, and another with exactly the same GWB signal realization plus an additional single source of continuous gravitational waves (CWs). 

The parameters of this simulation are chosen to be an interesting test, rather than a realistic possibility. For the GWB, we use an amplitude of $A_{\mathrm{gw}} = 3 \times  10^{-15}$ and a spectral index of  $\gamma_{\mathrm{gw}} = 13/3$. The CW was injected as a single face-on supermassive black-hole binary with a chirp mass of $\mathcal{M} = 1.1 \times 10^{9} M_\odot$, a distance of $D_\mathrm{CW}=100$~Mpc, and an Earth-term GW frequency corresponding exactly with the sixth frequency bin in the PTA, $f_{\rm CW}=1.46 \times 10^{-8} \, \mathrm{Hz}$. We also opt to exclude any binary evolution and pulsar-term contributions to ensure the the effects of the binary are constrained to this Earth-term frequency.

The PSD of timing residuals induced by a face-on supermassive black-hole binary---averaged over phase, polarization, and sky position---can be written as \citep{Ellis2012}
\begin{widetext}
\begin{equation}
    S_\mathrm{CW}(f_{\mathrm{CW}}) = 
    \frac{8\times10^{-15}}
    {12 \pi^2 \, f^2_{\mathrm{CW}}}
    \left(
    \frac{\mathcal{M}}{10^9 M_{\odot}}
    \right)^{10/3} 
    \left(
    \frac{100\,\mathrm{Mpc}}{D_{\mathrm{CW}}}
    \right)^{2} 
    \left(
    \frac{f_{\mathrm{CW}}}{50\,\mathrm{nHz}}
    \right)^{4/3}.
    \label{}
\end{equation}
\end{widetext}
%

With the aforementioned parameters, the combined PSD (i.e., GWB$+$CW) at the sixth frequency bin will be roughly five times larger than without the CW, without affecting any other frequencies. However, unlike the GWB, a CW's inter-pulsar correlation signature with a finite number of pulsars will only look approximately like the Hellings \& Downs curve when binned by angular separations, since some pulsars will be more affected than others based on their proximity to the GW source \citep{Cornish2013}. As such, the recovered PSD, which we model as Hellings \& Downs correlated, will only be an approximation, and may not match well in all cases.

\begin{figure}
    \centering
    \includegraphics[width=\columnwidth]{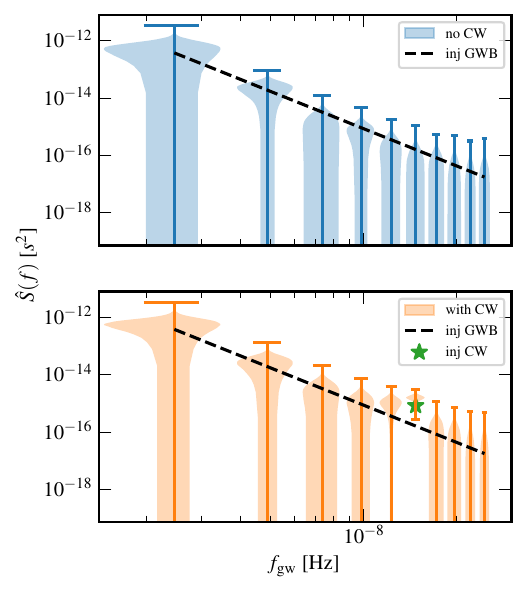}
    \caption{The recovered GWB spectrum from the PFOS using CURN sampling for parameter estimation with and without the CW present. The distributions on each frequency represent the uncertainty-sampled noise-marginalized distribution on the (broadband-normalized) PFOS' recovery. The dashed black line represents the injected GWB signal at an amplitude of $3 \times 10^{-15}$, and the green star represents the expected PSD value at the 6th frequency due to the single source.}
    \label{fig:PFOS_GWB+CW}
\end{figure}

For this test, we carry out an initial Bayesian power-law CURN search to provide our PTA parameter estimates, even when the CW was injected. This then mimics a situation in which the existence of the CW is unknown, thereby testing the ability for the PFOS to pick up on the excess PSD as a first-look discovery pipeline for single GW sources. \autoref{fig:PFOS_GWB+CW} shows the uncertainty-sampled and noise-marginalized PFOS (with pair covariance) PSD recovery for identical GWB simulations, with the top panel representing the GWB-only injection and the bottom panel containing both the GWB and CW injected at the sixth frequency. We see that even when the CW is unmodeled in the initial Bayesian PTA parameter estimation stage, the PFOS can pick up on the excess pulsar-pair correlated PSD caused by a CW. There are other differences to note however: most frequencies remain mostly unaffected by the CW (like those above the sixth frequency), some are modified slightly (like the first two frequencies), and the fifth frequency changes significantly along with the sixth frequency. We suspect that the reason for this is the now mis-estimated spectral shape in the pair-wise correlated measurements in \autoref{eq:rhof}.

Despite the differences, nearly all PSD recoveries are performing well. Every frequency in the GWB-only case includes the injected value within the 95\% credible region of the uncertainty sampled distributions. This remains the case for the GWB$+$CW injection for every frequency except the CW frequency. This frequency has the injected GWB+CW signal at just below the $3\sigma$ level; this isn't too surprising, as neglecting the contribution of the CW to the overall PSD leads to worse estimates of the total noise as well as the fact that the CW's correlations will not match those of the HD. 

While this test was mostly a proof of concept, we plan to further explore the ability of the PFOS to find the first CW sources in a future study.

\section{Discussion \& Conclusions}
\label{sec:discussion}


In this paper we presented a method to generalize the PTA optimal statistic to allow for a more general spectral model. We changed the Fourier-domain covariance matrix of the PTA from a model which assumes a power-law GWB to one which measures the GWB power spectrum at individual frequencies. The PFOS is capable of estimating the spectral shape of a correlated GWB in a fraction of the computational time of that of fully-correlated Bayesian PTA analyses; in fact the PFOS' computational time is dominated by the initial Bayesian common uncorrelated red noise search needed for noise and signal parameter estimates.  

In building the PFOS, we had to introduce an additional assumption of independent PSD estimates at each frequency across the PTA. Despite this slight mis-modeling, we showed in \autoref{subsec:per_frequency_recovery} that this assumption is valid for individual frequency measurements, and in \autoref{subsec:broadband_recovery} we showed that we can use these independent measurements to fit broadband spectral models, even in our realistic data sets. It is expected that with more pulsars, or more evenly-spaced timing measurements that are encountered in newer data sets \citep{NG15_dataset}, frequency covariance will become even less important.

Additionally, we found that in order to properly normalize the PFOS, it was important to introduce an estimate of the GWB spectral shape. This manifested as the unit-amplitude spectral shape at frequency $f$, $\Phi(f_k)$, that is analogous to the unit-power-law spectral shape, $\hat{\phi}$, in the traditional OS. We opted to treat this spectral shape estimate in a similar manner to the pulsar-pair covariant OS where this estimate is dictated by the CURN model parameters used to construct the (broadband-normalized) PFOS. \autoref{subsec:per_frequency_recovery} and \autoref{subsec:broadband_recovery} shows that using this strategy results in consistent measurements with injected values. We also introduced a separate form of the PFOS called the narrowband-normalized PFOS, which is entirely agnostic to the spectral shape. This benefit comes at the cost of introducing bias when individual pulsars contain frequency covariance in their measurements of their noise. \autoref{subsec:per_frequency_recovery} showed that despite this purposeful mis-modeling, the narrowband-normalized PFOS still delivers only slightly overestimated individual-frequency measurements. 
However, due to the systematic errors that are encountered when fitting a broad-spectrum model, it is preferred to use the far greater accuracy attained by the broadband-normalized PFOS.

We also touched on the choice of the Bayesian CURN search in \autoref{sec:PFOS_derivation} and \autoref{sec:sim_design}, where we found that using a variable spectral index power-law CURN model gave the best recovery, even in cases where there was an unmodeled single-source in the spectrum, as shown in \autoref{subsec:GWB+CW}. We also attempted to use models that are even more agnostic to the GWB spectral shape, such as modeling the CURN GWB PSD estimates independently; however, this resulted in significantly worse performance when these estimates were passed through to the broadband-normalized PFOS. This was likely due to the PSD estimates of the highest frequencies spanning multiple orders of magnitude, resulting in drastic changes to the spectral shape between the posterior draws passed to the PFOS during the noise marginalization stage. 

The PFOS represents the most general implementation of the PTA optimal statistic, relaxing the assumption of an assumed-shape GWB spectrum. In conjunction with other recently-developed methods like the multiple component optimal statistic \citep{Sardesai2023}, we now have the ability to characterize the structure of the spatial correlations and spectrum induced by a GWB at each GW frequency. There are several directions to continue development of the PFOS and to employ its flexibility. One may be able to avoid making any assumptions about the nature of the GWB spectrum (even for the purpose of providing an initial estimate) by restructuring the PFOS' $\chi^2$-fit to simultaneously model the contributions from all frequency bins. Initial work on this scheme shows some promise, though it is being plagued by numerical instability from the far larger number of elements in the pair covariance matrix (now $N_\mathrm{pairs}^2\times N_\mathrm{freq}^2$), and the correspondingly larger condition numbers. We plan to continue development of this approach in hopes of further strengthening the PFOS' accuracy. We also plan to continue exploring the effects of single GW sources on the correlated spectrum estimated by the PFOS. With the inclusion of anisotropic search techniques there may even be the potential for source localization \citep{NG15_anisotropy,pol2022anisotropy}. A key goal is to use the PFOS' capabilities to create a fast, all-frequency power search for continuous gravitational wave signals. This would act as a first-look statistic that may motivate, or act as proposal distributions for, subsequent template-based searches.

\section{Software}

The software used to implement these and other methods have been included in a comprehensive analysis python software package called \href{https://github.com/GersbachKa/defiant}{DEFIANT}. This software was made to incorporate the various expansions of the PTA optimal statistic, including the methods described in this paper, through modular, combinatory components. These components also include methods not featured including the multi-component optimal statistic \citep{Sardesai2023} and vastly improve on the performance of previous implementations with the inclusion of the pair covariance generalization \citep{Allen2023,Johnson2023}.

This software makes frequent use of the \href{https://github.com/nanograv/enterprise}{ENTERPRISE}\citep{enterprise} pulsar timing array software package for many of the needed matrix products and quantities.

\section{Acknowledgements}

The original coding implementation of the PTA optimal statistic's pair covariance was developed by Patrick Meyers and Michele Vallisneri, based on \citet{allen-romano2023}. We thank Holly Krynicki for many efficiency improvements in this code, which led the way for many other improvements made as part of this work. We thank William Lamb, Polina Petrov, and Levi Schult for cross-checks on both the mathematical and coding aspects of the techniques developed in this paper. We also thank Shashwat Sardesai for conversations relating to pulsar pair-covariance implementation. SRT and KAG acknowledge support from an NSF CAREER \#2146016. SRT also acknowledges support from NSF AST-2007993 and NSF AST-2307719.
JDR acknowledges suppport from start-up funds from the University of Texas Rio Grande Valley. The authors are members of the NANOGrav collaboration, which receives support from NSF Physics Frontiers Center award number 1430284 and
2020265. This work was conducted in part using the resources of the Advanced Computing Center for Research and Education (ACCRE) at Vanderbilt University, Nashville, TN.

\bibliography{pfos_bib}

\appendix
\renewcommand{\thesubsection}{\thesection\arabic{subsection}}

\section{Derivation of the Optimal Statistic's normalization}

In this appendix there are two derivations for the normalization. The first is for the traditional optimal statistic which assumes a power-law spectral template, and the second is for the per-frequency optimal statistic which is agnostic to the spectral template.

\subsection{Traditional Optimal Statistic}
\label{sec:os_normalization}

In order to calculate the normalization for \autoref{eq:os_pre_norm}, we set the expected value of our estimator to the desired quantity,
\begin{equation}
    \expect{ 
    \mathcal{N}_{ab} \, \bm{\delta t}_a^T \bm{P}_a^{-1} \hat{\bm{S}}_{ab} \bm{P}_b^{-1} \bm{\delta t}_b
    }=\Gamma_{ab} A^2_{\mathrm{gw}}
    .
    \label{eq:os_norm_start}
\end{equation}
To solve this equation, we rearrange the quantities and use two identities. First, the trace of a scalar is that same scalar, meaning that we can apply the trace to our estimator. Second, the expectation and trace are both linear operators that we can freely exchange. After rearranging and adding in the trace,
\begin{equation}
    \mathcal{N}_{ab} = \frac{\Gamma_{ab} A^2_{\mathrm{gw}}}
    { 
    \expect{\trace{
    \bm{\delta t}_a^T \bm{P}_a^{-1} \hat{\bm{S}}_{ab} \bm{P}_b^{-1} \bm{\delta t}_b}}
    }
    .
\end{equation}

From here, we use the cyclic property of the trace (i.e. with matrices $\bm{A}$, $\bm{B}$, and $\bm{C}$ we can cycle the elements, $\trace{\bm{A}\bm{B}\bm{C}}=\trace{\bm{B}\bm{C}\bm{A}}$). With this in mind we can cycle the trace and then exchange our trace and expectation value so that we apply the expectation to only the random variables in the equation, $\bm{\delta t}_b$ and $\bm{\delta t}_a^T$,
\begin{equation}
    \mathcal{N}_{ab} = \frac{\Gamma_{ab} A^2_{\mathrm{gw}}}
    {\trace{
    \bm{P}_b^{-1} \expect{\bm{\delta t}_b \, \bm{\delta t}_a^T} \bm{P}_a^{-1} \hat{\bm{S}}_{ab}
    }}
    .
\end{equation}

This allows us to substitute the expected covariance of TOA residuals, $\expect{\bm{\delta t}_b \; \bm{\delta t}_a^T} = \bm{S}_{ba} = \Gamma_{ab} A_{\mathrm{gw}}^2 \; \hat{\bm{S}}_{ba}$,
\begin{equation}
    \mathcal{N}_{ab} = \frac{\Gamma_{ab} A^2_{\mathrm{gw}}}
    {
    \Gamma_{ab} A_{\mathrm{gw}}^2 \trace{
    \bm{P}_b^{-1} \hat{\bm{S}}_{ba}
    \bm{P}_a^{-1} \hat{\bm{S}}_{ab}
    }}
    .
\end{equation}
Canceling the $\Gamma_{ab} A^2_{\mathrm{gw}}$ we find our final normalization
\begin{equation}
    \mathcal{N}_{ab} = \trace{
    \bm{P}_b^{-1} \hat{\bm{S}}_{ba} 
    \bm{P}_a^{-1} \hat{\bm{S}}_{ab}
    }^{-1}
    ,
\end{equation}
implying that our full cross-correlation estimator is then
\begin{equation}
    \rho_{ab} = \frac
    { \bm{\delta t}_a^T \bm{P}_a^{-1} \hat{\bm{S}}_{ab} \bm{P}_b^{-1} \bm{\delta t}_b}
    {\trace{ \bm{P}_b^{-1} \hat{\bm{S}}_{ba} \bm{P}_a^{-1} \hat{\bm{S}}_{ab}}}
    .
\end{equation}

\subsection{Per-Frequency Optimal Statistic}
\label{sec:PFOS_normalization}

The normalization calculation for the PFOS is similar to the traditional OS. However, we instead start from the rank-reduced form of the pair-wise estimator in \autoref{eq:pfos_pairwise_unnomed}. We set our expectation value so that
\begin{equation}
    \expect{ 
    \mathcal{N}_{ab}(f_k) \, \bm{X}_a^T  \Tilde{\phi}(f_k) \bm{X}_b
    }=\Gamma_{ab} \, S(f_k)
    .
\end{equation}
Substituting all $\bm{X}$ terms with $\bm{X}_a = \bm{F}^T_a \bm{P}_a^{-1} \bm{\delta t}_a$ then solving for $\mathcal{N}_{ab}(f_k)$ leads us to
\begin{equation}
    \mathcal{N}_{ab}(f_k) =
    \frac{\Gamma_{ab} \, S(f_k)}
    {\expect{ 
    \bm{\delta t}^T_a \bm{P}_a^{-1} \bm{F}_a \Tilde{\phi}(f_k) \bm{F}^T_b \bm{P}_b^{-1} \bm{\delta t}_b
    }}
    .
\end{equation}
Adding in the trace, we cycle it then exchange the trace with the expectation value to obtain
\begin{equation}
    \mathcal{N}_{ab}(f_k) = \frac{\Gamma_{ab} S(f_k)}
    { \trace{\bm{P}_a^{-1} \bm{F}_a \Tilde{\phi}(f_k) \bm{F}^T_b \bm{P}_b^{-1} \expect{\bm{\delta t}_b \; \bm{\delta t}^T_a}} }
    .
\end{equation}

Unlike with the traditional OS, we do not assume any particular spectral shape. Allowing for an arbitrary spectral shape, we write our expected cross-correlation as a spectrum, $\expect{\bm{\delta t}_b \; \bm{\delta t}^T_a} = \Gamma_{ab} \, \bm{F}_b \, \phi \, \bm{F}^T_a$. Substituting this into our equation then cycling the trace once again leads to
\begin{equation}
    \mathcal{N}_{ab}(f_k) = \frac{S(f_k)}
    { \trace{\bm{F}^T_a \bm{P}_a^{-1} \bm{F}_a \Tilde{\phi}(f_k) \bm{F}^T_b \bm{P}_b^{-1} \bm{F}_b \phi} }
    .
\end{equation}
Identifying our matrix substitution for $\bm{Z}_a = \bm{F}^T_a \bm{P}_a^{-1} \bm{F}_a$, we can simplify this equation as
\begin{equation}
    \mathcal{N}_{ab}(f_k) = \frac{S(f_k)}
    { \trace{ \bm{Z}_a \Tilde{\phi}(f_k) \bm{Z}_b \phi} }
    .
\end{equation}

As a final step, we ``normalize" our spectrum, meaning that we define a new quantity
\begin{equation}
    \Phi(f_k) = \frac{\phi}{S(f_k)}.
\end{equation}
Recall that $\phi$ is a diagonal matrix with elements corresponding to the Fourier modes of the GWB PSD, $\phi = \mathrm{diag}(S(f_1), S(f_1), S(f_2), S(f_2), ... , S(f_k), S(f_k), ...)$. Also recall that the shape of $\phi$ is ($2k_{\rm max} \times 2k_{\rm max}$) due to including both the sine and cosine of each frequency. This new quantity, $\Phi(f_k)$, represents the spectral shape of the GWB PSD with unit-value at frequency $f_k$. With this substitution we no longer need an estimate of $S(f_k)$; rather we need a spectral shape estimate. 

Substituting this normalization back into \autoref{eq:pfos_pairwise_unnomed}, we are left with our final estimator,
\begin{equation}
    \rho_{ab}(f_k) = \frac
    { \bm{X}^T_a \Tilde{\phi}(f_k) \bm{X}_b }
    { \trace{ \bm{Z}_a \Tilde{\phi}(f_k) \bm{Z}_b \Phi(f_k)} }
    .
\end{equation}

\vspace{10pt}
\section{Uncertainty on the OS cross-correlation estimators}

This appendix contains the derivations for the uncertainty in both the traditional optimal statistic and the per-frequency optimal statistic through the variance of the cross-correlation estimators. Both derivations share many of the same steps for the calculation, and are both calculated under the null hypothesis of an absence of pulsar-pair correlations. In these cases, the expectation values of the pair-wise correlated estimators are zero, as was assumed in the original formulation of the OS \citep{Anholm2009}. We will denote expectation values calculated assuming the null hypothesis by using a subscript $0$, so $\langle\ \rangle_0$.

\subsection{Traditional Optimal Statistic}
\label{sec:os_sigma_calc}

For the traditional optimal statistic, we have the cross-correlation estimator
\begin{equation}
    \rho_{ab} =
    \frac 
    {\bm{\delta t}_a^T \bm{P}_a^{-1}\hat{\bm{S}}_{ab}\bm{P}_b^{-1} \bm{\delta t}_b}
    {\trace{ \bm{P}_b^{-1} \hat{\bm{S}}_{ba}
    \bm{P}_a^{-1}\hat{\bm{S}}_{ab} }}
    .
\end{equation}
Calculating the variance of the OS cross-correlation estimator under the null hypothesis begins with recalling that the expectation of the individual pair-wise estimator is zero, so that
\begin{equation}
    \sigma_{ab,0}^2 = 
    \expect{\rho_{ab}^2}_0 - \expect{\rho_{ab}}^2_0 = 
    \expect{\rho_{ab}^2}_0
    .
\end{equation}
Substituting for $\rho_{ab}$ in our equation we find
\begin{equation}
    \sigma_{ab,0}^2 = 
    \frac{
    \expect{
    \bm{\delta t}_a^T \bm{P}_a^{-1} \hat{\bm{S}}_{ab}\bm{P}_b^{-1} \bm{\delta t}_b \,
    \bm{\delta t}_a^T \bm{P}_a^{-1} \hat{\bm{S}}_{ab} \bm{P}_b^{-1} \bm{\delta t}_b}_0
    }
    {
    \trace{ \bm{P}_b^{-1}\hat{\bm{S}}_{ba}
    \bm{P}_a^{-1}\hat{\bm{S}}_{ab} }^2
    }
    .
    \label{eq:expect_rho_square}
\end{equation}

Since we have four Gaussian random variables inside our expectation value, we must isolate these values from the constant matrix products. The first step involves using a temporary matrix product, $\bm{Q}_{ab} = \bm{P}_{a}^{-1} \: \hat{\bm{S}}_{ab} \: \bm{P}_b^{-1}$. We identify two instances of this quantity to create
\begin{equation}
    \sigma_{ab,0}^2 = 
    \frac{
    \expect{
    \bm{\delta t}^T_{a} \bm{Q}_{ab} \bm{\delta t}_{b}\,
    \bm{\delta t}^T_{a} \bm{Q}_{ab} \bm{\delta t}_{b}}_0
    }
    {
    \trace{ \bm{P}_b^{-1}\hat{\bm{S}}_{ba}
    \bm{P}_a^{-1}\hat{\bm{S}}_{ab} }^2
    }
    \label{eq:sigma_mid_calc}
\end{equation}

From here we move to index notation so that we can isolate our random variables. The indices $i$, $j$, $k$, and $l$ will represent the individual TOA residual measurements. This change results in
\begin{equation}
    \sigma_{ab,0}^2 = 
    \frac{ \sum_{ij,kl} \bm{Q}_{ab,ij} \, \bm{Q}_{ab,kl} 
    \expect{
    \bm{\delta t}_{a,i}\,\bm{\delta t}_{b,j}\,
    \bm{\delta t}_{a,k}\,\bm{\delta t}_{b,l}}_0 
    }
    {
    \trace{ \bm{P}_b^{-1} \hat{\bm{S}}_{ba}
    \bm{P}_a^{-1}\hat{\bm{S}}_{ab} }^2
    }
\end{equation}

We can evaluate the 4th order-moment of zero-mean Gaussian random variables using Isserlis' theorem~\cite{isserlis:1918}:
\begin{equation}
    \expect{ABCD} = \expect{AB}\expect{CD} + \expect{AC}\expect{BD} +\expect{AD}\expect{BC}.
\end{equation}
With this we write
\begin{widetext}
\begin{equation}
    \sigma_{ab,0}^2 = 
    \frac{ 
    \sum_{ij,kl} \bm{Q}_{ab,ij}\,\bm{Q}_{ab,kl} 
    }
    {
    \trace{ \bm{P}_b^{-1}\hat{\bm{S}}_{ba}
    \bm{P}_a^{-1}\hat{\bm{S}}_{ab} }^2
    }
    \left( 
    \expect{\bm{\delta t}_{a,i} \bm{\delta t}_{b,j}}_0 \expect{\bm{\delta t}_{a,k} \bm{\delta t}_{b,l}}_0 + 
    \expect{\bm{\delta t}_{a,i} \bm{\delta t}_{a,k}}_0 \expect{\bm{\delta t}_{b,j} \bm{\delta t}_{b,l}}_0 +
    \expect{\bm{\delta t}_{a,i} \bm{\delta t}_{b,l}}_0 \expect{\bm{\delta t}_{b,j} \bm{\delta t}_{a,k}}_0
    \right)
    .
\end{equation}
\end{widetext}

Next, we identify both the auto-correlation and cross-correlation matrices inside the expectation values. Since we are assuming the null hypothesis, we set to zero any cross-correlation quantities and identify auto-correlation matrices as $\bm{P}_{a,ij} = \expect{\bm{\delta t}_{a,i} \bm{\delta t}_{a,j}}_0$. This leaves us with
\begin{equation}
    \sigma_{ab,0}^2 = 
    \frac{ \sum_{ij,kl} \bm{Q}_{ab,ij} \, \bm{Q}_{ab,kl} \,
    \bm{P}_{a,ik} \, \bm{P}_{b,jl}}
    {\trace{ \bm{P}_b^{-1} \hat{\bm{S}}_{ba}
    \bm{P}_a^{-1} \hat{\bm{S}}_{ab} }^2}
    .
\end{equation}

Next, we use the fact that $\bm{P}_{a}$ is symmetric to write $\bm{P}_{a,ik} = \bm{P}_{a,ki}$. When combined with an additional matrix identity, $\bm{Q}_{ab,kl} = \bm{Q}_{ba,lk}$, we can rearrange the elements of the equation to be
\begin{equation}
    \sigma_{ab,0}^2 = 
    \frac{ \sum_{ij,kl}\bm{P}_{a,ki} \, \bm{Q}_{ab,ij} \, \bm{P}_{b,jl} \,\bm{Q}_{ba,lk} }
    {\trace{ \bm{P}_b^{-1} \hat{\bm{S}}_{ba}
    \bm{P}_a^{-1} \hat{\bm{S}}_{ab} }^2}
    .
\end{equation}

From here, we move back into matrix notation by identifying the trace of the matrix products in the numerator, letting us rewrite it as
\begin{equation}
    \sigma_{ab,0}^2 = 
    \frac{\trace{ \bm{P}_{a}\,\bm{Q}_{ab}\,\bm{P}_{b} \,\bm{Q}_{ba} }}
    {\trace{ \bm{P}_b^{-1} \hat{\bm{S}}_{ba}
    \bm{P}_a^{-1}\hat{\bm{S}}_{ab} }^2}
    .
    \label{eq:sigma_calc_near_end}
\end{equation}
Expressing our $\bm{Q}_{ab}$ quantities in terms of $\bm{P}^{-1}_a$ and $\hat{\bm{S}}_{ab}$, and canceling matrix products with their inverse, we are left with
\begin{equation}
    \sigma_{ab,0}^2 = 
    \frac{\trace{ \hat{\bm{S}}_{ab} \bm{P}_{b}^{-1}\hat{\bm{S}}_{ba} \bm{P}_a^{-1} }}
    {\trace{ \bm{P}_b^{-1} \hat{\bm{S}}_{ba}
    \bm{P}_a^{-1} \hat{\bm{S}}_{ab} }^2}
    ,
\end{equation}
which leads to the final (standard deviation) uncertainty for the cross-correlation estimators
\begin{equation}
    \sigma_{ab,0} = 
    \trace{ \bm{P}_b^{-1} \hat{\bm{S}}_{ba}
    \bm{P}_a^{-1}\hat{\bm{S}}_{ab} }^{-1/2}
    .
\end{equation}

\subsection{Per-Frequency Optimal Statistic}
\label{sec:pfos_uncertainty}

In this subsection, we calculate the uncertainty in the PFOS cross-correlated PSD estimator
\begin{equation}
    \rho_{ab}(f_k) = 
    \frac{ \bm{X}^T_a \Tilde{\phi}(f_k) \bm{X}_b }
    {\trace{ \bm{Z}_a \Tilde{\phi}(f_k) \bm{Z}_b \Phi(f_k)}} 
    .
\end{equation}
Much of the calculation is identical to Appendix \ref{sec:os_sigma_calc}, and as such many identical steps will be skipped. As was the case with the traditional OS, we calculate the uncertainty assuming the absence of a correlated-signal, such that the variance is then given by
\begin{equation}
    \sigma_{ab,0}(f_k)^2 = \expect{\rho_{ab,f}^2}_0 - \expect{\rho_{ab,f}}_0^2 
    = \expect{\rho_{ab,f}^2}_0
    .
\end{equation}

Substituting for $\rho_{ab}(f_k)$ leads to
\begin{equation}
    \sigma_{ab,0}(f_k)^2 = \expect{
    \frac{ \bm{X}^T_a \Tilde{\phi}(f_k) \bm{X}_b \; \bm{X}^T_a \Tilde{\phi}(f_k) \bm{X}_b}
    {\trace{ \bm{Z}_a \Tilde{\phi}_f \bm{Z}_b \Phi_f }^2}
    }_0
    ,
\end{equation}
We then substitute for $\bm{X}_a$ and $\bm{X}_b$ using $\bm{X}_a = \bm{F}^T_a \bm{P}_a^{-1} \bm{\delta t}_a$, and define a temporary matrix quantity $\bm{Q}'_{ab}(f_k) = \bm{P}_a^{-1} \bm{F}_a \Tilde{\phi}(f_k) \bm{F}_b^T \bm{P}_b^{-1}$. With this, we find
\begin{equation}
    \sigma_{ab,0}(f_k)^2 = 
    \frac{ \expect{ \bm{\delta t}^T_a \bm{Q}'_{ab}(f_k) \bm{\delta t}_b \, \bm{\delta t}^T_a \bm{Q}'_{ab}(f_k) \bm{\delta t}_b}_0 }
    {\trace{ \bm{Z}_a \Tilde{\phi}(f_k) \bm{Z}_b \Phi(f_k) }^2}
    .
\end{equation}

This equation has identical from to  \autoref{eq:sigma_mid_calc} from the traditional OS uncertainty calculation. If we follow through the same steps, we end up in a similar form as \autoref{eq:sigma_calc_near_end} except we now use rank-reduced quantities to obtain
\begin{equation}
    \sigma_{ab,0}(f_k)^2 =
    \frac{\trace{ \bm{Z}_a \Tilde{\phi}(f_k) \bm{Z}_b \Tilde{\phi}(f_k) } }
    {\trace{ \bm{Z}_a \Tilde{\phi}(f_k) \bm{Z}_b \Phi(f_k) }^2}
    .
\end{equation}
Finally, we take the square root to find the standard deviation uncertainty for the cross-correlated PSD at frequency $f_k$,
\begin{equation}
    \sigma_{ab,0}(f_k) = \left(
    \frac{\trace{ \bm{Z}_a \Tilde{\phi}(f_k) \bm{Z}_b \Tilde{\phi}(f_k) } }
    {\trace{ \bm{Z}_a \Tilde{\phi}(f_k) \bm{Z}_b \Phi(f_k) }^2}
    \right)^{1/2}
    .
\end{equation}

Unlike with the traditional OS, we cannot simplify this equation any further unless we make an additional assumption. One logical assumption, detailed in \autoref{sec:PFOS_derivation}, is that of the narrowband-normalized PFOS. This form of the PFOS assumes that each pulsar measures the GW frequencies independently. 
This leads us to replace $\Phi(f_k)$ with $\Tilde{\phi}(f_k)$, making these measurements independent of any assumed spectral shape. 
This lets us express the narrowband-normalized PFOS' uncertainty as
\begin{equation}
    \sigma_{ab,0,\mathrm{narrow}}(f_k) =
    \trace{ \bm{Z}_a \Tilde{\phi}(f_k) \bm{Z}_b \Tilde{\phi}(f_k) }^{-1/2}
    .
\end{equation}
While this may seem an attractive alternative, \autoref{subsec:per_frequency_recovery} and \autoref{subsec:broadband_recovery} detail how this additional simplification is not valid for realistic data. 

\section{Covariance of pairwise cross-correlation estimates} \label{sec:pair_covariance_calc}

As PTAs move towards larger signal regimes, the null-hypothesis assumption of zero covariance between pulsar pair estimators becomes a poor approximation, and the optimal statistic as a parameter estimator becomes severely biased, as was shown in \autoref{sec:par_est}. Including the pulsar-pair covariance fixes this problem. However, the number of elements in this covariance matrix scales as $N_{\mathrm{pulsars}}^4$. As such, this section will focus on deriving more efficient rank-reduced forms for calculating the elements of this covariance matrix for both the traditional OS and the PFOS.

\subsection{Traditional Optimal Statistic}
\label{sec:traditional_pair_covariance_calc}

With the traditional OS, we assume a power-law form for the GWB PSD. This assumption allows us to factor out a common GWB amplitude at a reference frequency of $f_\mathrm{yr}$ from our PSD matrix in \autoref{eq:powerlaw}, $\phi = A^2_{\mathrm{gw}} \, \hat{\phi}$, where $\hat{\phi}$ is the unit-amplitude power-law spectral template. 

The elements of the covariance matrix for the traditional OS can be represented as
\begin{equation}
\bm{C}_{ab,cd} = \expect{\rho_{ab}\, \rho_{cd}} - \expect{\rho_{ab}}\expect{\rho_{cd}}.
\end{equation}
For this calculation, we express our $\rho_{ab}$ quantities in a form that better translates to the PFOS case later. We write these in a similar manner to \autoref{eq:rho_ij} except we opt to factor the normalization and substitute $\hat{\bm{S}}_{ab} = \bm{F}_a \hat{\phi} \bm{F}^T_b$. With these changes, the estimator can now be written as
\begin{equation}
    \rho_{ab} = \mathcal{N}_{ab} \, 
    \bm{\delta t}^T_a \, \bm{P}^{-1}_a \, \bm{F}_a \, 
    \hat{\phi} \, 
    \bm{F}^T_b \, \bm{P}^{-1}_b \, \bm{\delta t}_b
    ,
\end{equation}
where $\mathcal{N}_{ab} = \trace{\bm{Z}_a \hat{\phi} \bm{Z}_b \hat{\phi}}^{-1}$ is the normalization for pulsar pair $ab$, computed in Appendix \ref{sec:os_normalization}. For convenience, we now define a temporary matrix quantity
\begin{equation}
    \bm{Q}_{ab} = \bm{P}^{-1}_a \, \bm{F}_a \, \hat{\phi} \, \bm{F}^T_b \, \bm{P}^{-1}_b
    ,
\end{equation}
which allows us to rewrite our estimator as
\begin{equation}
    \rho_{ab} = \mathcal{N}_{ab} \, 
    \bm{\delta t}^T_a \, \bm{Q}_{ab} \, \bm{\delta t}_a
    .
\end{equation}

We now substitute this into our covariance matrix, letting us factor out our normalization to find
\begin{equation}
\begin{aligned}
    \bm{C}_{ab,cd} = 
    \mathcal{N}_{ab} \: \mathcal{N}_{cd} \big[ & \expect{
    \bm{\delta t}_a^T \, \bm{Q}_{ab} \, \bm{\delta t}_b \;
    \bm{\delta t}_c^T \, \bm{Q}_{cd} \, \bm{\delta t}_d 
    }  \\ & - 
    \expect{\bm{\delta t}_a^T \, \bm{Q}_{ab} \, \bm{\delta t}_b}
    \expect{\bm{\delta t}_c^T \, \bm{Q}_{cd} \, \bm{\delta t}_d}
    \big)
    .
\label{eq:pair_cov_first_step}
\end{aligned}
\end{equation}

To solve this, we will work term by term. Solving for the first term in this equation requires much of the same process as was done in Appendix \ref{sec:os_sigma_calc} and requires us to move to index notation,
\begin{equation}
\begin{aligned}
    &\expect{ \bm{\delta t}_a^T \, \bm{Q}_{ab} \, \bm{\delta t}_b \;
    \bm{\delta t}_c^T \, \bm{Q}_{cd} \, \bm{\delta t}_d } = \\
    & \hspace{2cm} \sum_{ijkl} 
    \bm{Q}_{ab,ij} \, \bm{Q}_{cd,kl} \expect{
    \bm{\delta t}_{a,i}  \, \bm{\delta t}_{b,j} \,
    \bm{\delta t}_{c,k}  \, \bm{\delta t}_{d,l} 
    }
    .
\end{aligned}
\end{equation}

We now apply relationships for the 4$^{\mathrm{th}}$ moment for our zero mean, Gaussian random variables \citep{isserlis:1918}. This will turn our equation into the expectation of pairs of TOA residuals which we simplify to cross-correlations, $\expect{\bm{\delta t}_{a,i} \, \bm{\delta t}_{b,j}} = \bm{S}_{ab,ij}$. Applying these steps, we end up with
\begin{widetext}
\begin{equation}
    \expect{ \bm{\delta t}_a^T \, \bm{Q}_{ab} \, \bm{\delta t}_b \;
    \bm{\delta t}_c^T \, \bm{Q}_{cd} \, \bm{\delta t}_d } 
    = \sum_{ijkl}  \bm{Q}_{ab,ij} \, \bm{Q}_{cd,kl} \\
    \left[ 
    \bm{S}_{ab,ij}\,\bm{S}_{cd,kl}+
    \bm{S}_{ac,ik}\,\bm{S}_{bd,jl}+
    \bm{S}_{ad,il}\,\bm{S}_{bc,jk}
    \right]
    .
\end{equation}
\end{widetext}

We can then distribute the $\bm{Q}$ quantities, and apply the following identities, $\bm{S}_{ab,ij} = \bm{S}_{ba,ji}$, and $\bm{Q}_{ab,ij} = \bm{Q}_{ba,ji}$, to find traces of matrix products. The resulting set of traces create
\begin{equation}
\begin{aligned}
    \expect{ \bm{\delta t}_a^T \, \bm{Q}_{ab} \, \bm{\delta t}_b \;
    \bm{\delta t}_c^T \, \bm{Q}_{cd} \, \bm{\delta t}_d } 
    & = \trace{\bm{S}_{ab}\bm{Q}_{ba}} \trace{\bm{S}_{cd}\bm{Q}_{dc}} \\ 
    & +
    \trace{\bm{S}_{ac}\bm{Q}_{cd}\bm{S}_{db}\bm{Q}_{ba}} \\
    & + \trace{\bm{S}_{ad}\bm{Q}_{dc}\bm{S}_{cb}\bm{Q}_{ba}}
    .
\end{aligned}
\end{equation}

Focusing now on the second term from \autoref{eq:pair_cov_first_step}, we can use the same methodology as we did in \autoref{sec:os_normalization}. After adding traces, we cycle the matrices to then evaluate the expectation values,
\begin{equation}
    \expect{\bm{\delta t}_a^T \, \bm{Q}_{ab} \, \bm{\delta t}_b} 
    \expect{\bm{\delta t}_c^T \, \bm{Q}_{cd} \, \bm{\delta t}_d} =
    \trace{\bm{S}_{ba} \, \bm{Q}_{ab}} 
    \trace{\bm{S}_{dc} \, \bm{Q}_{cd}}
    .
\end{equation}
Note that this term cancels out part of the first, leaving us with a more compact representation,
\begin{equation}
\begin{aligned}
    \bm{C}_{ab,cd} = 
    \mathcal{N}_{ab} \: \mathcal{N}_{cd} \big( 
    & \trace{\bm{S}_{ac}\bm{Q}_{cd}\bm{S}_{db}\bm{Q}_{ba}}  \\ & + 
    \trace{\bm{S}_{ad}\bm{Q}_{dc}\bm{S}_{cb}\bm{Q}_{ba} }
    \big) 
    .
\label{eq:pair_covariance_non_reduced}
\end{aligned}
\end{equation}

In this form, this equation is the same as the one found in Equation~$69$ of Section~IV.B of \citet{Johnson2023}. However, implementing pulsar-pair covariance in this form will be computationally expensive due to the $\bm{Q}_{ab}$ matrix being $(N_{\mathrm{TOA},a} \times N_{\mathrm{TOA},b})$. Instead we can adopt the $\bm{X}$ and $\bm{Z}$ rank-reduced matrices to compress our matrix representations. It also becomes more helpful to identify the unique cases of $\bm{C}_{ab,cd}$. These three cases are: $0$ matching pulsars ($\bm{C}_{ab,cd}$), $1$ matching pulsar ($\bm{C}_{ab,ac}$), and $2$ matching pulsars ($\bm{C}_{ab,ab}$).

\subsubsection{0-match case}

For the $0$-match case, $\bm{C}_{ab,cd}$, there are no simplifications that we can make from \autoref{eq:pair_covariance_non_reduced} before substituting for the temporary $\bm{Q}_{ab}$ matrices and the cross-correlation matrices $\bm{S}_{ab} = A^2_{\mathrm{gw}} \Gamma_{ab} \, \bm{F}_a \hat{\phi} \bm{F}^T_b$. This leaves us with
\begin{widetext}
\begin{equation}
\begin{aligned}
    \bm{C}_{ab,cd} = 
    \mathcal{N}_{ab} \: \mathcal{N}_{cd} \bigg( 
    & A^4_{\mathrm{gw}} \, \Gamma_{ac} \Gamma_{bd} \, \trace{\bm{F}_a \hat{\phi} \bm{F}^T_c \bm{P}^{-1}_c \bm{F}_c \hat{\phi} \bm{F}^T_d \bm{P}^{-1}_d
    \bm{F}_d \hat{\phi} \bm{F}^T_b \bm{P}^{-1}_b \bm{F}_b \hat{\phi} \bm{F}^T_a  \bm{P}^{-1}_a
    }  \\ & + 
    A^4_{\mathrm{gw}} \, \Gamma_{ad} \Gamma_{bc} \, \trace{\bm{F}_a \hat{\phi} \bm{F}^T_d \bm{P}^{-1}_d \bm{F}_d \hat{\phi} \bm{F}^T_c \bm{P}^{-1}_c 
    \bm{F}_c \hat{\phi} \bm{F}^T_b \bm{P}^{-1}_b \bm{F}_b \hat{\phi} \bm{F}^T_a \bm{P}^{-1}_a }
    \bigg)
    .
\end{aligned}
\end{equation}
In this form, we can cycle our matrices inside the trace and identify the rank-reduced matrix product $\bm{Z}_a = \bm{F}^T_a \bm{P}^{-1}_a \bm{F}_a$. This lets us simplify our equation into
\begin{equation}
    \bm{C}_{ab,cd} = 
    \mathcal{N}_{ab} \, \mathcal{N}_{cd} \bigg(
    A^4_{\mathrm{gw}} \, \Gamma_{ac} \Gamma_{bd} \, 
    \trace{ \bm{Z}_a \hat{\phi} \bm{Z}_c \hat{\phi} 
            \bm{Z}_d \hat{\phi} \bm{Z}_b \hat{\phi} } 
    + A^4_{\mathrm{gw}} \, \Gamma_{ad} \Gamma_{bc} \, \trace{\bm{Z}_a \hat{\phi} \bm{Z}_d \hat{\phi} \bm{Z}_c \hat{\phi} \bm{Z}_b \hat{\phi} }
    \bigg)
    .
\label{eq:pc_nomatch}
\end{equation}

\end{widetext}

\subsubsection{1-match case}

For the $1$-match case, $\bm{C}_{ab,ac}$, we can make an additional simplification to \autoref{eq:pair_covariance_non_reduced} by replacing $\bm{S}_{aa}$ with $\bm{P}_a$, with $\bm{P}_a$ potentially containing a contribution from the GWB. This leaves us with a slightly compressed version,
\begin{equation}
\begin{aligned}
    \bm{C}_{ab,ac} = 
    \mathcal{N}_{ab} \, \mathcal{N}_{ac} \big( 
    & \trace{\bm{P}_{a}\bm{Q}_{ac}\bm{S}_{cb}\bm{Q}_{ba}}  \\ & + 
    \trace{\bm{S}_{ac}\bm{Q}_{ca}\bm{S}_{ab}\bm{Q}_{ba} }
    \big) 
    .
\label{eq:pc_onematch}
\end{aligned}
\end{equation}

Following the same strategy as the $0$-match case, we substitute for the $\bm{Q}_{ab}$ and $\bm{S}_{ab}$ matrices, cycle the traces, and identify the $\bm{Z}_a$ quantities. Additionally we simplify $\bm{P}_a \bm{P}^{-1}_a = I$. This leaves
\begin{equation}
\begin{aligned}
    \bm{C}_{ab,ac} = 
    \mathcal{N}_{ab} \, \mathcal{N}_{ac} \bigg( 
    & A^2_{\mathrm{gw}} \, \Gamma_{cb} \trace{ \bm{Z}_{b} \hat{\phi} \bm{Z}_{a} \hat{\phi} \bm{Z}_{c} \hat{\phi} } \\ & + 
    A_{\mathrm{gw}}^4 \, \Gamma_{ac} \Gamma_{ab} \trace{ 
    \bm{Z}_{b} \hat{\phi} \bm{Z}_{a} \hat{\phi} \bm{Z}_{c} \hat{\phi} \bm{Z}_{a} \hat{\phi}}
    \bigg) 
    .
\label{eq:pc_twomatch}
\end{aligned}
\end{equation}

Note that, despite appearing as though there are four unique $1$-match scenarios ($\bm{C}_{ab,ac}$, $\bm{C}_{ab,bc}$, $\bm{C}_{ab,ca}$, and $\bm{C}_{ab,cb}$), they are all mathematically equivalent through index swapping and transpose identities.

\subsubsection{2-match case}

Finally, in the $2$-match case, $\bm{C}_{ab,ab}$, we further simplify \autoref{eq:pair_covariance_non_reduced} to
\begin{equation}
\begin{aligned}
    \bm{C}_{ab,ab} = 
    \mathcal{N}^2_{ab} \big( 
    & \trace{\bm{P}_{a}\bm{Q}_{cd}\bm{P}_{b}\bm{Q}_{ba}} 
     \\ & + 
    \trace{\bm{S}_{ab}\bm{Q}_{dc}\bm{S}_{ab}\bm{Q}_{ba} }
    \big)
    .
\end{aligned}
\end{equation}
Again, applying the strategies to rank-reduce, we find more products of matrices with their inverses which allows us to simplify to,
\begin{equation}
\begin{aligned}
    \bm{C}_{ab,ab} = 
    \mathcal{N}^2_{ab} \bigg( 
    & \trace{\bm{Z}_{b} \hat{\phi} \bm{Z}_{a} \hat{\phi}} 
     \\ & + 
    \Gamma_{ab}^2 \, A_{\mathrm{gw}}^4 
    \trace{ \bm{Z}_{b} \hat{\phi} \bm{Z}_{a} \hat{\phi} \bm{Z}_{b} \hat{\phi} \bm{Z}_{a} \hat{\phi}}
    \bigg) 
    .
\end{aligned}
\end{equation}

\subsubsection{Equivalence with the traditional optimal statistic}

Interestingly, we can also see why the traditional OS is effective in the weak signal regime through these calculations. If one assumes that the effect of the GWB is negligible, then we can set $A^2_{\mathrm{gw}} = 0$. This zeroes the covariance matrix for all off-diagonals and leaves the remaining diagonal to be
\begin{equation}
    \bm{C}_{ab,ab} = 
    \mathcal{N}^2_{ab} \left( 
    \trace{\bm{Z}_{b} \hat{\phi} \bm{Z}_{a} \hat{\phi}} 
    \right) = 
    \trace{\bm{Z}_{b} \hat{\phi} \bm{Z}_{a} \hat{\phi}}^{-1}
    ,
\end{equation}
which is equivalent to the variance of the traditional OS in \autoref{eq:c_traditional}.

\subsection{Per-frequency optimal statistic}

Deriving pair-covariance with the PFOS is nearly identical to that of the traditional OS, with the largest difference being the $\phi$ quantities. We start by writing our covariance matrix as
\begin{equation}
    \bm{C}_{ab,cd} = \expect{\rho_{ab}(f_k) \rho_{cd}(f_k)} - \expect{\rho_{ab}(f_k)} \expect{\rho_{cd}(f_k)}
    .
\end{equation}
We then define our per-frequency cross-correlation estimators as
\begin{equation}
    \rho_{ab}(f_k) = \mathcal{N}_{ab}(f_k) \bm{\delta t}^T_a \bm{P}_a^{-1} \bm{F}_a \Tilde{\phi}(f_k) \bm{F}^T_b \bm{P}^{-1}_b \bm{\delta t}_b
    ,
\end{equation}
where $\mathcal{N}_{ab}(f_k) = \trace{\bm{Z}_a \Tilde{\phi}(f_k) \bm{Z}_b \Phi(f_k)}^{-1}$. If we then use a temporary matrix quantity $\bm{Q}_{ab}(f_k) = \bm{P}_a^{-1} \bm{F}_a \Tilde{\phi}(f_k) \bm{F}^T_b \bm{P}^{-1}_b$, we can now write our estimator in a format identical to the traditional OS case,
\begin{equation}
    \rho_{ab}(f_k) = \mathcal{N}_{ab}(f_k) \bm{\delta t}^T_a \bm{Q}_{ab}(f_k) \bm{\delta t}_b
    .
\end{equation}

With this format, we use the same strategies and methods to solve for the covariance in each case. The difference come from the way we define $\phi$. As with the PFOS, we decompose $\bm{S}_{ab} = S(f_k) \, \Gamma_{ab} \,\bm{F}_a \Phi(f_k) \bm{F}^T_b$, allowing for a variable spectral shape. Running through the same processes, we get expressions for the following cases.

\begin{widetext}
\subsubsection{0-match case}
\begin{equation} 
\begin{aligned}
    \bm{C}_{ab,cd}(f_k) = \mathcal{N}_{ab}(f_k) \,\mathcal{N}_{cd}(f_k) \bigg( &
    \Gamma_{ac} \Gamma_{bd} \, S(f_k)^2 \, \trace{ 
    \bm{Z}_{b} \Tilde{\phi}(f_k) \bm{Z}_{a} \Phi(f_k) \bm{Z}_{c} \Tilde{\phi}(f_k) \bm{Z}_{d} \Phi(f_k)} 
     \\ & + 
    \Gamma_{ad} \Gamma_{bc} \, S(f_k)^2 \, \trace{ 
    \bm{Z}_{b} \Tilde{\phi}(f_k) \bm{Z}_{a} \Phi(f_k) \bm{Z}_{d} \Tilde{\phi}(f_k) \bm{Z}_{c} \Phi(f_k)} 
    \bigg) 
\end{aligned}
\end{equation}

\subsubsection{1-match case}
\begin{equation}
\begin{aligned}
    \bm{C}_{ab,ac}(f_k) = \mathcal{N}_{ab}(f_k) \, \mathcal{N}_{ac}(f_k) \bigg(
    & \Gamma_{bc} \, S(f_k) \, \trace{ 
    \bm{Z}_{b} \Tilde{\phi}(f_k) \bm{Z}_{a} \Tilde{\phi}(f_k) \bm{Z}_{c} \Phi(f_k)} 
     \\ & + 
    \Gamma_{ac} \Gamma_{ab} \, S(f_k)^2 \, \trace{ 
    \bm{Z}_{b} \Tilde{\phi}(f_k) \bm{Z}_{a} \Phi(f_k) \bm{Z}_{c} \Tilde{\phi}(f_k) \bm{Z}_{a} \Phi(f_k)} 
    \bigg) 
\end{aligned}
\end{equation}

\subsubsection{2-match case}
\begin{equation}
\begin{aligned}
    \bm{C}_{ab,ab}(f_k) = \mathcal{N}_{ab}(f_k)^2 
    \bigg(
    & \trace{ 
    \bm{Z}_{b} \Tilde{\phi}(f_k) \bm{Z}_{a} \Tilde{\phi}(f_k) } 
    \\ & + 
    \Gamma_{ab}^2 \, S(f_k)^2 \, \trace{ 
    \bm{Z}_{b} \Tilde{\phi}(f_k) \bm{Z}_{a} \Phi(f_k) \bm{Z}_{b} \Tilde{\phi}(f_k) \bm{Z}_{a} \Phi(f_k)} 
    \bigg)
\end{aligned}
\end{equation}
\end{widetext}

\end{document}